\documentclass[3p, sort&compress, a4paper]{elsarticle}

\usepackage[]{graphicx}
\usepackage{url}
\usepackage{ifpdf}
\usepackage{hyperref}
\hypersetup{breaklinks,colorlinks,citecolor=blue}
\usepackage{amssymb}
\usepackage{amsmath}
\usepackage{mathtools}
\DeclarePairedDelimiter{\ceil}{\lceil}{\rceil}
\DeclarePairedDelimiter{\floor}{\lfloor}{\rfloor}
\usepackage{subfigure}
\usepackage{color}

\newcommand{\eqn}[1]{(#1)}
\newcommand{\Eqn}[1]{(#1)}

\newcommand{\fig}[1]{Fig.~#1}

\newcommand{\sectn}[1]{Sec.~#1}

\newcommand{\eg}{\mbox{\it e.g.}}
\newcommand{\ie}{\mbox{\it i.e.}}

\newcommand{\cf}{\mbox{\it cf.}}

\newcommand{\cmb}{{CMB}}
\newcommand{\cmbtext}{{cosmic microwave background}}

\newcommand{\healpix}{{\tt HEALPix}}

\newcommand{\sshtcode}{{\tt SSHT}}
\newcommand{\sothreecode}{{\tt SO3}}
\newcommand{\stwoletcode}{{\tt S2LET}}

\newcommand{\spcend}{\ensuremath{\:}}
\newcommand{\img}{\ensuremath{{\rm i}}}
\newcommand{\cconj}{\ensuremath{\ast}} 
 
\newcommand{\reals}{\ensuremath{\mathbb{R}}}

\newcommand{\realsnz}{\ensuremath{\mathbb{R}^{+}_{\ast}}}
\newcommand{\integers}{\ensuremath{\mathbb{Z}}}
\newcommand{\naturals}{\ensuremath{\mathbb{N}}}

\newcommand{\ltwo}{\ensuremath{\mathrm{L}^2}}
\newcommand{\sphere}{\ensuremath{{\mathbb{S}^2}}}
\newcommand{\sothree}{\ensuremath{{\mathrm{SO}(3)}}}

\newcommand{\vect}[1]{\ensuremath{\mbox{\boldmath ${#1}$}}}

\newcommand{\dx}{\ensuremath{\mathrm{\,d}}}
\newcommand{\dmu}[1]{\ensuremath{\dx \Omega(#1)}}

\newcommand{\deul}[1]{\ensuremath{\dx \varrho(#1)}}

\newcommand{\innerp}[2]{\ensuremath{\langle {#1},\: {#2} \rangle}}

\newcommand{\sa}{\ensuremath{\omega}}
\newcommand{\saa}{\ensuremath{\theta}}
\newcommand{\sab}{\ensuremath{\varphi}}
\newcommand{\sas}{\ensuremath{\saa, \sab}}
\newcommand{\eul}{\ensuremath{\mathbf{\rho}}}
\newcommand{\euls}{\ensuremath{\eula, \eulb, \eulc}}
\newcommand{\eula}{\ensuremath{\alpha}}
\newcommand{\eulb}{\ensuremath{\beta}}
\newcommand{\eulc}{\ensuremath{\gamma}}

\newcommand{\eulci}{\ensuremath{g}}

\newcommand{\eulciang}{\ensuremath{\eulc_\eulci}}
\newcommand{\el}{\ensuremath{\ell}}
\newcommand{\m}{\ensuremath{m}}
\newcommand{\n}{\ensuremath{n}}

\newcommand{\spin}{\ensuremath{s}}

\newcommand{\elmax}{\ensuremath{{L}}}
\newcommand{\mmax}{\ensuremath{{M}}}
\newcommand{\nmax}{\ensuremath{{N}}}

\newcommand{\p}{\ensuremath{^\prime}}

\newcommand{\kron}[2]{\ensuremath{\delta_{{#1}{#2}}}}

\renewcommand{\exp}[1]{\ensuremath{{\rm e}^{#1}}}
\newcommand{\shfarg}[3]{\ensuremath{Y_{#1#2}({#3})}}

\newcommand{\sshfarg}[4]{\ensuremath{{{}_{#4} Y_{#1#2}({#3})}}}
\newcommand{\sshfargc}[4]{\ensuremath{{{}_{#4} Y_{#1#2}^\cconj({#3})}}}
\newcommand{\sshfargsp}[4]{\ensuremath{{{}_{#4} Y_{{#1},{#2}}({#3})}}}
\newcommand{\shf}[2]{\ensuremath{Y_{#1#2}}}

\newcommand{\shc}[3]{\ensuremath{{#1}_{{#2}{#3}}}}
\newcommand{\shcc}[3]{\ensuremath{{#1}_{{#2}{#3}}^\cconj}}
\newcommand{\shcsp}[3]{\ensuremath{{#1}_{{#2},{#3}}}}

\newcommand{\sshf}[3]{\ensuremath{{{}_{#3} Y_{#1#2}}}}

\newcommand{\sshc}[4]{\ensuremath{{}_{#4} {#1}_{{#2}{#3}}}}

\newcommand{\dmatbig}{\ensuremath{D}}
\newcommand{\Dlmn}{\ensuremath{ \dmatbig_{\m\n}^{\el} }}
\newcommand{\Dlmnc}{\ensuremath{ \dmatbig_{\m\n}^{\el\cconj} }}

\newcommand{\Dlmnpc}{\ensuremath{ \dmatbig_{\m\n}^{\el\cconj}(\eul) }}

\newcommand{\wigc}[4]{\ensuremath{{#1}^{#2}_{{#3}{#4}}}}

\newcommand{\rot}{\ensuremath{\mathcal{R}}}
\newcommand{\rotarg}[1]{\ensuremath{\mathcal{R}_{#1}}}

\newcommand{\rotmatarg}[1]{\ensuremath{\mathbf{R}_{#1}}}

\newcommand{\spinup}{\ensuremath{\eth}}
\newcommand{\spindown}{\ensuremath{\bar{\eth}}}

\newcommand{\f}{\ensuremath{f}}
\newcommand{\fs}{\ensuremath{{}_\spin f}}
\newcommand{\fsm}{\ensuremath{{{}_{-\spin} f}}}

\newcommand{\fslm}{\ensuremath{\shc{\fs}{\el}{\m}}}

\newcommand{\wav}{\ensuremath{\psi}}
\newcommand{\swav}{\ensuremath{{}_\spin\psi}}

\newcommand{\wavs}{\ensuremath{\Phi}}
\newcommand{\swavs}{\ensuremath{{}_\spin\Phi}}
\newcommand{\wcoeff}{\ensuremath{W}}
\newcommand{\scoeff}{\ensuremath{W}}
\newcommand{\wscale}{\ensuremath{j}}
\newcommand{\wscalemax}{\ensuremath{J}}
\newcommand{\wscalemin}{\ensuremath{J_0}}

\newcommand{\dilparam}{\ensuremath{\alpha}}
\newcommand{\wavker}{\ensuremath{\kappa}}
\newcommand{\wavsteer}{\ensuremath{s}}
\newcommand{\steerinterp}{\ensuremath{z}}

\newcommand{\sumlm}{\ensuremath{\sum_{\el=0}^{\infty} \sum_{\m=-\el}^\el}}
\newcommand{\sumlmn}{\ensuremath{\sum_{\el=0}^{\infty} \sum_{\m=-\el}^\el} \sum_{\n=-\el}^\el}
\newcommand{\suml}{\ensuremath{\sum_{\el=0}^{\infty}}}
\newcommand{\sumlmbl}{\ensuremath{\sum_{\el=0}^{\elmax-1} \sum_{\m=-\el}^\el}}

\newcommand{\summ}{\ensuremath{\sum_{\m=-\el}^\el}}

\newcommand{\sumn}{\ensuremath{\sum_{\n=-\el}^\el}}

\newcommand{\stokesi}{\ensuremath{I}}
\newcommand{\stokesq}{\ensuremath{Q}}
\newcommand{\stokesu}{\ensuremath{U}}
\newcommand{\stokesv}{\ensuremath{V}}
\newcommand{\emode}{\ensuremath{E}}
\newcommand{\bmode}{\ensuremath{B}}
\newcommand{\emodetilde}{\ensuremath{\tilde{E}}}
\newcommand{\bmodetilde}{\ensuremath{\tilde{B}}}
\newcommand{\qpmiu}{\ensuremath{\stokesq \pm \img \stokesu}}
\newcommand{\qpiu}{\ensuremath{\stokesq + \img \stokesu}}
\newcommand{\qmiu}{\ensuremath{\stokesq - \img \stokesu}}

\newcommand{\order}{\ensuremath{\mathcal{O}}}

\newcommand{\convdir}{\ensuremath{\circledast}}
\newcommand{\convaxisym}{\ensuremath{\odot}}

\renewcommand{\eqn}[1]{Eq.~(#1)}
\renewcommand{\Eqn}[1]{Eq.~(#1)}
\renewcommand{\wav}{\ensuremath{\Psi}}
\renewcommand{\swav}{\ensuremath{{}_\spin\Psi}}
\renewcommand{\wavsteer}{\ensuremath{\zeta}}
\renewcommand{\dilparam}{\ensuremath{{\lambda}}}

\renewcommand{\rotmatarg}[1]{\ensuremath{\mathsf{R}_{#1}}}
\renewcommand{\mmax}{\ensuremath{{N}}}

\renewcommand{\sshtcode}{{\sc ssht}}
\renewcommand{\sothreecode}{{\sc so3}}
\renewcommand{\stwoletcode}{{\sc s2let}}
\renewcommand{\healpix}{{\sc healpix}}
\newcommand{\fftwcode}{{\sc fftw}}

\begin{document}


\begin{frontmatter}

\title{Directional spin wavelets on the sphere}

\author[mssl]{Jason~D.~McEwen\fnref{fn1}}
\ead{jason.mcewen@ucl.ac.uk}

\author[panda]{Boris~Leistedt\fnref{fn2}}

\author[panda]{Martin~B\"uttner\fnref{fn3}}

\author[panda]{Hiranya~V.~Peiris\fnref{fn4}}

\author[hwu]{Yves~Wiaux}

\address[mssl]{Mullard Space Science Laboratory (MSSL), University College
  London (UCL), Surrey RH5 6NT, UK}

\address[panda]{Department of Physics and Astronomy, University College
  London (UCL), London WC1E UK}

\address[hwu]{Institute of Sensors, Signals, and Systems, Heriot-Watt
  University, Edinburgh EH14 4AS, UK}

\fntext[fn1]{Partially supported by the Engineering and Physical Sciences
Research Council (grant number EP/M011852/1).}

\fntext[fn2]{Partially supported by the Impact and Perren funds and by the
European Research Council under the European Community's Seventh Framework
Programme (FP7/2007-2013) / ERC grant agreement no 306478-CosmicDawn.}

\fntext[fn3]{Supported by New Frontiers in Astronomy and Cosmology grant \#37426
and by the Royal Astronomical Society.}

\fntext[fn4]{Partially supported by the European Research Council under the
European Community's Seventh Framework Programme (FP7/2007-2013) / ERC grant
agreement no 306478-CosmicDawn.}


\begin{abstract}
  We construct a directional spin wavelet framework on the sphere by
  generalising the scalar scale-discretised wavelet transform to signals of
  arbitrary spin.  The resulting framework is the only wavelet framework defined
  natively on the sphere that is able to probe the directional intensity of spin
  signals.  Furthermore, directional spin scale-discretised wavelets support the
  exact synthesis of a signal on the sphere from its wavelet coefficients and
  satisfy excellent localisation and uncorrelation properties.  Consequently,
  directional spin scale-discretised wavelets are likely to be of use in a wide
  range of applications and in particular for the analysis of the polarisation
  of the \cmbtext\ (\cmb).  We develop new algorithms to compute (scalar and
  spin) forward and inverse wavelet transforms exactly and efficiently for very
  large data-sets containing tens of millions of samples on the sphere.  By
  leveraging a novel sampling theorem on the rotation group developed in a
  companion article, only half as many wavelet coefficients as alternative
  approaches need be computed, while still capturing the full information
  content of the signal under analysis.  Our implementation of these algorithms
  is made publicly available.
\end{abstract}


\end{frontmatter}

\section{Introduction}

Wavelet transforms on the sphere \cite[\eg][]{schroder:1995, sweldens:1997,
antoine:1999, antoine:1998, barreiro:2000, wiaux:2005, sanz:2006,
mcewen:2006:cswt2, starck:2006, mcewen:2008:fsi, mcewen:szip, narcowich:2006,
baldi:2009, marinucci:2008, wiaux:2007:sdw, leistedt:s2let_axisym,
mcewen:2013:waveletsxv, chan:s2let_curvelets, mcewen:s2let_ridgelets} to analyse
scalar signals defined on a spherical domain have found widespread use.  For
example, wavelet analyses on the sphere have led to many insightful scientific
studies in planetary science \cite[\eg][]{audet:2010}, geophysics 
\cite[\eg][]{simons:2011} and cosmology, in particular for the analysis of the
\cmbtext\ (\cmb) 
\cite[\eg][]{vielva:2004, mcewen:2005:ng, mcewen:2006:ng,
  mcewen:2008:ng, mcewen:2006:bianchi, vielva:2005, mcewen:2006:isw,
  mcewen:2007:isw2, vielva:2006, wiaux:2008, wiaux:2006, lan:2008,
  pietrobon:2006, planck2013-p06, planck2013-p09, 
  planck2013-p20, schmitt:2011, bobin:2013, delabrouille:2009, rogers:s2let_ilc_temp, rogers:s2let_ilc_pol, leistedt:ebsep};  
for a review see \cite{mcewen:2006:review}.

Discrete wavelet frameworks on the sphere that can support the exact
synthesis of signals from their wavelet coefficients in a stable
manner have received a great deal of attention recently, including:
needlets \cite{narcowich:2006, baldi:2009, marinucci:2008};
directional scale-discretised wavelets \cite{wiaux:2007:sdw,
  leistedt:s2let_axisym, mcewen:2013:waveletsxv}; and the isotropic
undecimated and pyramidal wavelet transforms \cite{starck:2006}.  All
three of these approaches have also been extended to the three-dimensional (3D) ball formed by augmenting the sphere
with the radial line \cite{durastanti:2014, leistedt:flaglets,
  mcewen:flaglets_sampta, lanusse:2012}, such as the distribution of
galaxies in our Universe. 

Data acquired on spherical domains come in a variety of different
forms: not only scalar but also vector and tensor signals defined on
the sphere.  Spin functions \cite{newman:1966} provide a unifying
framework for representing such signals, where spin $\spin=0$
functions are scalar, $\spin = \pm 1$ functions are vector, and higher
spin orders correspond to tensor functions.  An example of a spin
$\spin = \pm 1$ signal of interest is geomagnetic anomalies over the
Earth \cite{plattner:2013}, while an example of a spin $\spin = \pm 2$
signal is the polarised radiation of the \cmb\
\cite{zaldarriaga:1997, kamionkowski:1996}.  Motivated largely by the analysis of \cmb\
polarisation, spin wavelet transforms on the sphere have been
developed recently \cite{starck:2009, geller:2008, geller:2010:sw,
  geller:2010}, focusing predominantly on the spin
$\spin = \pm 2$ setting.

The isotropic undecimated and pyramidal wavelet transforms on the
sphere \cite{starck:2006} have been extended to spin $\spin = \pm 2$
signals \cite{starck:2009}.  A number of constructions are proposed in
\cite{starck:2009}, settling on a transform constructed by firstly
decomposing a spin-2 signal into its parity even and odd components,
so-called E- and B-mode signals \cite{zaldarriaga:1997, kamionkowski:1996}, before
performing the scalar wavelet transform of \cite{starck:2006}.  This
construction does not easily generalise to signals of arbitrary spin
and is restricted to wavelets that can only probe isotropic signal
content.  The spin curvelet construction of \cite{starck:2009} that
supports a directional analysis is built on the twelve base-resolution
faces of the \healpix\ \cite{gorski:2005} pixelisation, as are the
scalar curvelets of \cite{starck:2009}, and as such is not natively
defined on the sphere.  This approach leads to (blocking) artefacts \cite{starck:2006}.

Needlets have also been extended to the analysis of spin functions on
the sphere, in a variety of manners \cite{geller:2008, geller:2010:sw,
  geller:2010}.  The standard spin needlet construction
\cite{geller:2008, geller:2010:sw} is defined by a linear projector
onto the space of spin functions.  Consequently, the needlet
coefficients of spin signals are spin quantities themselves, of the
same spin number.  Alternatively, the so-called mixed needlet
construction \cite{geller:2010} results in spin needlet coefficients
that are scalar quantities.  Both spin needlet constructions satisfy
excellent localisation and uncorrelation properties and generalise
easily to signals of arbitrary spin number \cite{geller:2008,
  geller:2010:sw, geller:2010}.  However, spin needlet constructions
can also only probe isotropic signal content.

A spin wavelet framework on the sphere capable of probing the
directional intensity of spin signals does not yet exist.  We develop
such a framework in this article, generalising directional
scale-discretised wavelets \cite{wiaux:2007:sdw,
  leistedt:s2let_axisym, mcewen:2013:waveletsxv} to the analysis of
signals on the sphere of arbitrary spin. 
Directionality provides an additional source of localisation, where
oriented signal features can be probed.  For example, it is widely known that
the peaks of the CMB, in models where the CMB is assumed to be a realisation
of an isotropic Gaussian random field, are predicted to be elongated
\cite{bond:1987}.  Indeed, directional analyses of CMB temperature data have revealed signatures
of physical processes and effects that have been detected at high statistical
significance \cite[\eg][]{mcewen:2005:ng,mcewen:2006:isw,mcewen:2007:isw2}. 
In related articles, we construct spin ridgelet
\cite{mcewen:s2let_ridgelets} and curvelet \cite{chan:s2let_curvelets}
transforms defined natively on the sphere and built on the general
spin scale-discretised wavelet framework presented herein.  In another
recent article \cite{mcewen:s2let_localisation} we show that
directional scale-discretised wavelets satisfy excellent localisation
and uncorrelation problems similar to needlets, in both the scalar
setting and the spin generalisation presented here.

An overview of our spin scale-discretised wavelet construction was
first presented in \cite{mcewen:s2let_spin_sccc21_2014}. In
this article, however, we present a more complete construction, prove a number
of important properties of the construction, develop new algorithms to
compute the wavelet transform exactly and efficiently, evaluate our
implementation of these algorithms, and present an illustrative
application.
The remainder of this article is structured as follows.  Harmonic
analysis on the sphere and rotation group is reviewed concisely in
\sectn{\ref{sec:harmonic_analysis}}.  The directional spin
scale-discretised wavelet framework is derived in
\sectn{\ref{sec:spin_wavelets}}, while an algorithm to compute the
wavelet transform and its inverse exactly and efficiently for very
large data-sets is presented and evaluated in
\sectn{\ref{sec:computation}}. Applications are discussed and an illustrative denoinsing example is
presented in \sectn{\ref{sec:applications}}, before concluding remarks
are made in \sectn{\ref{sec:conclusions}}.

\section{Harmonic analysis on the sphere and rotation group}
\label{sec:harmonic_analysis}

We concisely review harmonic analysis on the sphere and rotation group
in this section, presenting the mathematical preliminaries required
throughout the remainder of the article.  General spin signals on the
sphere are first reviewed, before we specialise to spin-2 signals and
highlight alternative representations of practical interest.
Signals on the rotation group are then discussed, followed by a
description of the rotation of functions on the sphere, expressed in
both the spatial and harmonic domains.

\subsection{Spin signals on the sphere}

Square integrable spin functions on the sphere
$\fs\in \ltwo(\sphere)$, with integer spin $\spin\in\integers$, are
defined by their behaviour under local rotations.  By definition, a
spin function transforms as
\cite{newman:1966,goldberg:1967,zaldarriaga:1997,kamionkowski:1996}
\begin{equation}
  \label{eqn:spin_rot}
  \fs^\prime(\sa) = \exp{-\img \spin \chi} \: \fs(\sa)
\end{equation}
under a local rotation by $\chi \in [0,2\pi)$, where the prime denotes
the rotated function.\footnote{The sign convention adopted for the
  argument of the complex exponential 
  differs to the original definition \cite{newman:1966} but is
  identical to the convention used in the context of the polarisation
  of the \cmb\ \cite{zaldarriaga:1997,kamionkowski:1996}.  }  It is important to note
that the rotation considered here is \emph{not} a global rotation on
the sphere
but rather a rotation by $\chi$ in the tangent plane centred on the
spherical coordinates $\sa=(\sas)$, with colatitude $\saa \in [0,\pi]$ and
longitude $\sab \in [0,2\pi)$.

The spin spherical harmonics $\sshf{\el}{\m}{\spin} \in \ltwo(\sphere)$ form an
orthogonal basis for $\ltwo(\sphere)$ spin \spin\ functions on the
sphere, for natural $\el\in\naturals$ and integer $\m\in\integers$,
$|\m|\leq\el$, $|\spin|\leq\el$.  We adopt the Condon-Shortley phase
convention, such that the conjugate symmetry relation
\mbox{$\sshfargc{\el}{\m}{\sas}{\spin} = (-1)^{\spin+\m}
  \sshfargsp{\el}{-\m}{\sas}{-\spin}$} holds.
The orthogonality and completeness relations for the spherical
harmonics read 
\begin{equation}
\innerp{\sshf{\el}{\m}{\spin}}
{\sshf{\el\p}{\m\p}{\spin}}
 = 
\kron{\el}{\el\p}
\kron{\m}{\m\p}
\end{equation}
and
\begin{equation}
  \sumlm
  \sshfarg{\el}{\m}{\saa,\sab}{\spin} \:
  \sshfargc{\el}{\m}{\saa\p,\sab\p}{\spin}
  =
  \delta(\cos\saa - \cos\saa\p) \:
  \delta(\sab - \sab\p),
\end{equation}
respectively, 
where
$\kron{i}{j}$ is the Kronecker delta symbol and $\delta(\cdot)$ is the
Dirac delta function.
The inner product of $f,g\in\ltwo(\sphere)$ is defined by
\begin{equation}
  \innerp{f}{g} = \int_\sphere \dmu{\sa} \: f(\sa) \: g^\cconj(\sa) 
  \spcend ,
\end{equation}
where $\dmu{\sas} = \sin \saa \dx \saa \dx \sab$ is the usual
invariant measure on the sphere and complex conjugation is denoted by
${\cdot}^\cconj$.

Spin raising and lowering operators, $\spinup$ and $\spindown$
respectively, increment and decrement the spin order of a spin-\spin\
function by unity and are defined by
\begin{equation}  
  \spinup \equiv
  -\sin^\spin \saa 
  \Bigl ( 
  \frac{\partial}{\partial \saa} 
  + \frac{\img}{\sin\saa} \frac{\partial}{\partial \sab}
  \Bigr)
  \sin^{-\spin}\saa
\end{equation}
and
\begin{equation}  
  \spindown \equiv 
  -\sin^{-\spin} \saa 
  \Bigl ( 
  \frac{\partial}{\partial \saa} 
  - \frac{\img}{\sin\saa} \frac{\partial}{\partial \sab}
  \Bigr)
  \sin^{\spin}\saa
  \spcend ,
\end{equation}
respectively \cite{newman:1966,goldberg:1967,zaldarriaga:1997,kamionkowski:1996}.  
The spin-\spin\ spherical harmonics can thus be expressed in terms of
the scalar (spin-zero) harmonics through the spin raising and lowering
operators by \cite{newman:1966,goldberg:1967,zaldarriaga:1997,kamionkowski:1996}
\begin{equation}  
  \sshfarg{\el}{\m}{\sa}{\spin} 
  =
  \biggl[ \frac{(\el-\spin)!}{(\el+\spin)!} \biggr]^{1/2} 
  \spinup^\spin
  \shfarg{\el}{\m}{\sa}
  \spcend ,
\end{equation}
for $0 \leq \spin \leq \el$,
and by 
\begin{equation}  
  \sshfarg{\el}{\m}{\sa}{\spin} 
  =
  (-1)^\spin
  \biggl[ \frac{(\el+\spin)!}{(\el-\spin)!} \biggr]^{1/2} 
  \spindown^{-\spin}
  \shfarg{\el}{\m}{\sa}
  \spcend ,
\end{equation}
for $-\el \leq \spin \leq 0$, where $\shf{\el}{\m}$ denote the scalar
(spin-zero) spherical harmonics.

Due to the orthogonality and completeness of the spin spherical
harmonics, any square integrable spin function on the sphere
$\fs \in \ltwo(\sphere)$ may be represented by its spherical harmonic
expansion
\begin{equation}
\fs(\sa) = 
\sum_{\el=0}^\infty
\sum_{\m=-\el}^\el
\fslm \:
\sshfarg{\el}{\m}{\sa}{\spin}
\spcend ,
\end{equation}
where the spin spherical harmonic coefficients are given by the usual
projection onto each basis function: $\fslm =
\innerp{\fs}{\sshf{\el}{\m}{\spin}}$.
The conjugate symmetry relation of the spin spherical harmonic
coefficients is given by
$\shcc{\fs}{\el}{\m} = (-1)^{\spin+\m} \:
{}_{-\spin}\shcsp{\f}{\el}{-\m}$
for a function satisfying \mbox{$\fs^\cconj=\fsm$} (which for a spin
$\spin=0$ function equates to the usual reality condition) and follows
directly from the conjugate symmetry of the spin spherical harmonics.
Throughout, we consider signals on the sphere band-limited at
$\elmax$, that is signals such that $\shc{\fs}{\el}{\m}=0$,
$\forall \el\geq\elmax$.
The spherical harmonic transform can be computed exactly
and efficiently for band-limited signals by appealing to sampling
theorems on the sphere and fast algorithms \cite{driscoll:1994, wiaux:2005b,mcewen:fsht, mcewen:fssht, mcewen:so3}.

\subsection{Spin-2 signals on the sphere}
\label{sec:harmonic_analysis:spin2}

Although we typically consider signals of arbitrary spin throughout,
we specialise to spin-2 signals on the sphere in some cases to
illustrate connections between scalar and spin-2 wavelet transforms.
Moreover, the spin-2 setting is of particular interest for
the analysis of linearly polarised radiation.  We briefly review
spin-2 signals in this context.

The Stokes parameters of polarised radiation are denoted
$\stokesi,\stokesq,\stokesu,\stokesv \in \ltwo(\sphere)$, where
$\stokesi$ encodes the intensity, $\stokesq$ and $\stokesu$ the linear
polarisation of the incident radiation, and $\stokesv$ the circular
polarisation component, which is assumed to be zero.  The linear
polarisation signal that is observed depends on the choice of local
coordinate frame.  The component $(\qpmiu)$ transforms under a
rotation of the local coordinate frame by $\chi \in [0, 2\pi)$ as
$(\qpmiu)^\prime(\sa) = \exp{( \mp \img 2 \chi)}(\qpmiu)(\sa)$ and is
thus a spin $\pm 2$ signal on the sphere \cite{zaldarriaga:1997,kamionkowski:1996}
(hereafter the appropriate spin number subscript is denoted).  The
quantity ${}_{\pm2}(\qpmiu)$ can be decomposed into parity even and
odd components, so-called E- and B-modes, by
\begin{equation}
  \label{eqn:e_modes}
  \emodetilde(\sa) = 
  -\frac{1}{2}
  \bigl[ \spindown^2 {}_{2} (\qpiu)(\sa) 
  + {\spinup}^2 {}_{-2} (\qmiu)(\sa) \bigr] 
\end{equation}
and
\begin{equation}
  \label{eqn:b_modes}
  \bmodetilde(\sa) = 
  \frac{\img}{2}
  \bigl[ \spindown^2 {}_{2} (\qpiu)(\sa) 
  - {\spinup}^2 {}_{-2} (\qmiu)(\sa) \bigr] 
  \spcend ,
\end{equation}
respectively, where $\emodetilde,\bmodetilde \in \ltwo(\sphere)$ are
scalar signals.  The spin spherical harmonic
coefficients of the ${}_{\pm2}(\qpmiu)$ Stokes signal are related to
the scalar spherical harmonic coefficients of the E- and B-mode
signals by
\begin{equation}
  \sshc{(\qpmiu)}{\el}{\m}{\pm 2} =
  -(
  \shc{\emode}{\el}{\m}
  \pm \img
  \shc{\bmode}{\el}{\m} )
  \spcend ,
\end{equation}
where
$\shc{\bmodetilde}{\el}{\m} = \sqrt{\frac{(\el+2)!}{(\el-2)!}} \: \shc{\bmode}{\el}{\m}$
and
$\shc{\emodetilde}{\el}{\m} = \sqrt{\frac{(\el+2)!}{(\el-2)!}} \: \shc{\emode}{\el}{\m}$.

\subsection{Signals on the rotation group}

Since we construct directional wavelet transforms, we also consider
square integrable functions on the rotation group
\mbox{$f \in \ltwo(\sothree)$}, where rotations are parameterised by
the Euler angles $\eul=(\euls)$, with
$\eula \in [0,2\pi)$, $\eulb \in [0,\pi]$ and $\eulc \in [0,2\pi)$.
We adopt the $zyz$ Euler convention corresponding to the rotation of a
physical body in a \emph{fixed} coordinate system about the $z$, $y$
and $z$ axes by $\eulc$, $\eulb$ and $\eula$, respectively.

The Wigner $\dmatbig$-functions $\Dlmn \in \ltwo(\sothree)$, with natural $\el\in\naturals$
and integer $\m,\n\in\integers$, $|\m|,|\n|\leq\el$, are the matrix
elements of the irreducible unitary representation of the rotation
group \sothree\  \cite{varshalovich:1989}.  Consequently, the \Dlmnc\ also form an orthogonal
basis in $\ltwo(\sothree)$.\footnote{We adopt the conjugate
  \dmatbig-functions as basis elements since this convention
  simplifies connections to wavelet transforms on the sphere.}  
The Wigner $\dmatbig$-functions satisfy the conjugate symmetry
relation
$\dmatbig_{\m\n}^{\el\cconj}(\eul) = (-1)^{\m+\n}
\dmatbig_{-\m,-\n}^{\el}(\eul)$.
The
orthogonality and completeness relations for the Wigner
$\dmatbig$-functions read
$
\innerp{\dmatbig_{\m\n}^{\el}}{\dmatbig_{\m\p\n\p}^{\el\p}}
= 
8\pi^2
\kron{\el}{\el\p}
\kron{\m}{\m\p}
\kron{\n}{\n\p} / (2\el+1)
$
and 
\begin{equation}
  \suml \summ \sumn 
  \dmatbig_{\m\n}^{\el}(\eula,\eulb,\eulc) \:
  \dmatbig_{\m\n}^{\el\cconj}(\eula\p,\eulb\p,\eulc\p) 
   =
  \delta(\eula - \eula\p)
  \delta(\cos\eulb - \cos\eulb\p) \:
  \delta(\eulc - \eulc\p)
  \spcend ,
\end{equation}
respectively \cite{varshalovich:1989}.
The inner product of $f,g\in\ltwo(\sothree)$ is defined by
$
  \innerp{f}{g} = \int_\sothree \deul{\eul} \: f(\eul) \: g^\cconj(\eul) 
$,  
where $\deul{\eul} = \sin\eulb \dx\eula \dx\eulb \dx\eulc$ is the
usual invariant measure on the rotation group.  Note that
$\innerp{\cdot}{\cdot}$ is used to denote inner products over both the
sphere and the rotation group (the case adopted can be inferred from
the context).
The Wigner \dmatbig-functions may also be related to the spin spherical
harmonics by \cite{goldberg:1967}
\begin{equation}
  \label{eqn:ssh_wigner}
  \exp{-\img \spin \eulc}
  \sshfarg{\el}{\m}{\eulb,\eulc}{\spin} = (-1)^\spin 
  \sqrt{\frac{2\el+1}{4\pi} } \:
  \dmatbig_{\m,-\spin}^{\el\:\cconj}(\euls)
  \spcend .
\end{equation}

Due to the orthogonality and completeness of the Wigner
$\dmatbig$-functions, any square integrable function on the rotation
group $\f \in \ltwo(\sothree)$ may be represented by its Wigner
expansion
\begin{equation}
  \label{eqn:wig_inverse}
   \f(\eul) = 
   \sum_{\el=0}^{\infty} \frac{2\el+1}{8\pi^2} \sum_{\m=\el}^{\el}
   \sum_{\n=-\el}^\el
   \wigc{\f}{\el}{\m}{\n} \Dlmnpc
   \spcend,
\end{equation}
where the Wigner coefficients are given by the projection onto each
basis function: $\wigc{\f}{\el}{\m}{\n} = \innerp{\f}{\Dlmnc}$.
The conjugate symmetry relation of the Wigner
coefficients is given by
$\wigc{\f}{\el \: \cconj}{\m}{\n}  = (-1)^{\m+\n} \:
\wigc{\f}{\el}{-\m,}{-\n}$
for a real function satisfying \mbox{$\f^\cconj=\f$}  and follows
directly from the conjugate symmetry of the Wigner $\dmatbig$-functions.
Throughout, we consider signals on the rotation group band-limited at
$\elmax$, that is signals such that $\wigc{\f}{\el}{\m}{\n}=0$,
$\forall \el\geq\elmax$.  The Wigner transform can be computed exactly
and efficiently for band-limited signals by appealing to sampling
theorems on the rotation group and fast algorithms \cite{kostelec:2008,
  mcewen:so3}.

\subsection{Rotation of signals on the sphere }

We define the rotation of a spin function on the sphere \mbox{$\fs \in \ltwo(\sphere)$}
by
\begin{equation}
  \label{eqn:spin_rotation}
  (\rotarg{\eul} \: \fs)(\sa) 
  \equiv 
  \exp{-\img \spin \vartheta}
  \fs(\rotmatarg{\eul}^{-1}\vect{\hat{\sa}})
  \spcend ,
\end{equation}
where $\vect{\hat{\sa}} \in \reals^3$ denotes the Cartesian vector
corresponding to $\sa$ and $\rotmatarg{\eul}$ is the 3D rotation matrix
corresponding to the rotation defined by $\eul=(\euls)$.  The angle $\vartheta
\in [0, 2\pi)$ is defined by the third Euler angle of the rotation
$\rotmatarg{\eul\p} = \rotmatarg{\eul}^{-1} \rotmatarg{\sa}$, \ie\
\mbox{$\eul\p=(\cdot,\cdot,\vartheta)$}, where $\rotmatarg{\sa}$ is the 3D rotation
matrix corresponding to a rotation defined by $(\euls)=(\sab,\saa,0)$ for
$\sa=(\sas)$.  The exponential factor appearing in
\eqn{\ref{eqn:spin_rotation}} is required to ensure that the rotation of a
spin $\spin$ function results in a function with the same spin order.%
\footnote{In general, performing purely a coordinate rotation of a spin
function does not result in a rotated function with the same spin number. This
can be seen by considering the special property of spin functions about the
poles.  At the pole, the tangent coordinate frame coincides with the azimuthal
coordinate $\sab$.  Hence, the spin symmetry relation (\eg\ invariance under
a rotation by $\pi$ for a spin $\spin=\pm2$ function) must also hold in $\sab$
at the poles.  Although this symmetry  holds only locally at the poles, since
band-limited functions are smooth, it also holds approximately in the vicinity
of the poles.  Any rotation of a spin function where the poles rotate to a
point that is not a pole will break this symmetry property since there is no
guarantee that the rotated function will exhibit the required symmetry at the
poles.  Consequently, the rotated function will not be of the same spin as the
original function.}
The magnitude of a spin function is rotated in the usual manner (\ie\ through
a coordinate rotation), however the additional phase factor means that
the real and imaginary components of the signal are not rotated solely by a
coordinate rotation.  For the scalar setting where $\spin=0$,
\eqn{\ref{eqn:spin_rotation}} reduces to the typical rotation operator
defined solely through a rotation of the coordinate system.

It is convenient to also express the rotation operator in harmonic space.
First, we note the additive property of the Wigner \dmatbig-functions
given by \cite{marinucci:2011:book}
\begin{equation}
  \dmatbig_{\m\n}^{\el}(\eul) 
  =
  \sum_{k=-\el}^\el
  \dmatbig_{\m k}^{\el}(\eul_1) \:
  \dmatbig_{k \n}^{\el}(\eul_2) 
  \spcend ,
\end{equation}
where $\eul$ describes the rotation formed by composing the
rotations described by $\eul_1$ and $\eul_2$, \ie\
$\rotmatarg{\eul} = \rotmatarg{\eul_1} \rotmatarg{\eul_2}$.  By noting
the relation between the Wigner \dmatbig-functions and the spin
spherical harmonics of \eqn{\ref{eqn:ssh_wigner}}, it follows that the
spin spherical harmonics can be rotated by
\begin{equation}
  \label{eqn:spherical_harmonic_rotation}
  (\rotarg{\eul} \: \sshf{\el}{\m}{\spin})(\sa)
  =
  \sum_{\n=-\el}^\el
  \dmatbig_{\n \m}^{\el}(\eul) \:
  \sshfarg{\el}{\n}{\sa}{\spin}
  \spcend ,
\end{equation}
and thus 
\begin{equation}
  \label{eqn:spherical_harmonic_coeff_rotation}
  \shc{(\rotarg{\eul} \: \fs)}{\el}{\m}
  =
  \sum_{\n=-\el}^\el
  \dmatbig_{\m \n}^{\el}(\eul) \:
  \sshc{f}{\el}{\n}{\spin}
  \spcend .
\end{equation}
From \eqn{\ref{eqn:spherical_harmonic_rotation}} it is apparent that the
rotation of a spin $\spin$ spherical harmonic, with rotation operator defined
by \eqn{\ref{eqn:spin_rotation}}, results in a function of the same spin
number $\spin$.  Consequently, this property also holds for general spin
functions: with rotation operator defined by
\eqn{\ref{eqn:spin_rotation}}, the rotation of a spin function results in a
function with the same spin number.

\section{Directional spin wavelet transform on the sphere}
\label{sec:spin_wavelets}

The directional spin scale-discretised wavelet transform on the sphere
is developed in this section.  We present the wavelet analysis and
synthesis of a signal of arbitrary spin (\ie\ forward and inverse
wavelet transforms), before describing the construction of the
wavelets themselves.  Important properties of the directional spin
wavelet transform are then discussed.

\subsection{Wavelet analysis}
\label{sec:spin_wavelets:analysis}

The spin scale-discretised wavelet transform on the sphere of a spin
function $\fs \in \ltwo(\sphere)$ may be defined analogously to the
scalar transform, with wavelet coefficients given by the directional
convolution
\begin{equation}
  \wcoeff^{\swav^{(\wscale)}}(\eul) 
  \equiv ( \fs \convdir \swav^{(\wscale)}) (\eul)
  \equiv \innerp{\fs}{\rotarg{\eul} \: \swav^{(\wscale)}}
  = \int_\sphere \dmu{\sa} \: \fs(\sa) \: (\rotarg{\eul} \: \swav^{(\wscale)})^\cconj(\sa)
  \label{eqn:wav_analysis}
  \spcend ,
\end{equation}
where the operator $\convdir$ denotes directional convolution on the
sphere and the wavelet $\swav^{(\wscale)} \in \ltwo(\sphere)$ is now
also a spin function on the sphere.
The wavelets are designed to be localised in scale, position and
orientation, and are constructed explicitly in
\sectn{\ref{sec:spin_wavelets:construction}}.  The wavelet transform
of \eqn{\ref{eqn:wav_analysis}} thus probes directional structure in
the signal of interest \fs, where \eulc\ of $\eul=(\euls)$ can be
viewed as the orientation about each point on the sphere
$(\sas) = (\eulb, \eula)$.  The wavelet scale
$\wscale \in \naturals_0$ encodes the angular localisation of
$\wav^{(\wscale)}$.

By decomposing the function \fs\ and wavelet $\swav^{(\wscale)}$ into
their spherical harmonic expansions, and noting the orthogonality
of the spin spherical harmonics and their rotation by
\eqn{\ref{eqn:spherical_harmonic_rotation}}, the wavelet transform of
\eqn{\ref{eqn:wav_analysis}} may be written
\begin{equation}
  \label{eqn:wav_analysis_harmonic}
  \wcoeff^{\swav^{(\wscale)}}(\eul)   
  = \sumlmn
  \sshc{\f}{\el}{\m}{\spin} \:
  \sshc{\wav}{\el}{\n}{\spin}^{(\wscale)\cconj} \:
  \Dlmnpc  
  \spcend,
\end{equation}
where $\fslm = \innerp{\fs}{\sshf{\el}{\m}{\spin}}$ and
$\sshc{\wav}{\el}{\m}{\spin}^{(\wscale)} =
\innerp{\swav^{(\wscale)}}{\sshf{\el}{\m}{\spin}}$
are the spin spherical harmonic coefficients of the function of
interest and wavelet,
respectively. \Eqn{\ref{eqn:wav_analysis_harmonic}} is the spin
generalisation of the harmonic representation of the directional
convolution typically considered for scalar functions
\cite{mcewen:2006:fcswt, mcewen:2013:waveletsxv, wiaux:2005,
  wiaux:2005c, wiaux:2007:sdw}.
From \eqn{\ref{eqn:wav_analysis_harmonic}} it is apparent that the
spin wavelet transform results in scalar wavelet coefficients defined
on the rotation group.
Comparing \eqn{\ref{eqn:wav_analysis_harmonic}} and
\eqn{\ref{eqn:wig_inverse}} it is apparent that the Wigner
coefficients of the wavelet coefficients defined on \sothree\ are
given by
\begin{equation}
  \label{eqn:wav_harmonic_coeff}
  \wigc{\bigl(\wcoeff^{\swav^{(\wscale)}}\bigr)}{\el}{\m}{\n}
  = \frac{8 \pi^2}{2\el+1} \:
  \sshc{\f}{\el}{\m}{\spin} \:
  \sshc{\wav}{\el}{\n}{\spin}^{(\wscale)\cconj}
  \spcend ,
\end{equation}
where
$\wigc{\bigl(\wcoeff^{\swav^{(\wscale)}}\bigr)}{\el}{\m}{\n} =
\innerp{ \wcoeff^{\swav^{(\wscale)}}}{\Dlmnc}$,
as noted previously for the scalar setting
\cite{mcewen:2013:waveletsxv, wiaux:2005, wiaux:2007:sdw}.  As we
elaborate further in \sectn{\ref{sec:computation:algorithm}}, the
forward wavelet transform of \eqn{\ref{eqn:wav_analysis}} may thus be computed
via an inverse Wigner transform.

The wavelets do not probe the low-frequency content of the signal \fs;
hence, a scaling function \mbox{$\swavs \in \ltwo(\sphere)$} is
introduced for this purpose, with scaling coefficients
\mbox{$\wcoeff^{\swavs^{(\wscale)}} \in \ltwo(\sphere)$} given by the
axisymmetric convolution 
\begin{equation}
  \label{eqn:analysis_scaling}
  \wcoeff^{\swavs}(\sa)\equiv ( \fs \convaxisym \swavs) (\sa)
  \equiv \innerp{\fs}{\rotarg{\sa}\:\swavs}
  = \int_\sphere \dmu{\sa\p} \:\fs(\sa\p) \:
  (\rotarg{\sa}\:\swavs)^\cconj(\sa\p)
  \spcend ,
\end{equation}
where we adopt the shorthand notation
$\rotarg{\sa} = \rotarg{(\sab,\saa,0)}$ and the operator $\convaxisym$
denotes axisymmetric convolution on the sphere with kernels that are
invariant under azimuthal rotations when centred on the North pole.
We design the scaling function to be axisymmetric such that
$\rotarg{(0,0,\eulc)} \swavs =\swavs$ since directional structure of
the low-frequency signal content probed by the scaling function is not
typically of interest.  Consequently, the spin spherical harmonic
coefficients of the scaling function are non-zero for $\m=0$ only:
$\sshc{\wavs}{\el}{0}{\spin} \kron{\m}{0} =
\innerp{\swavs}{\sshf{\el}{\m}{\spin}}$.

It follows from \eqn{\ref{eqn:wav_analysis_harmonic}} that the scaling
coefficients may be decomposed into their harmonic expansion
\begin{equation}
  \label{eqn:analysis_scaling_harmonic}
  \wcoeff^{\swavs}(\sa)   
  = \sumlm
  \sqrt{\frac{4 \pi}{2 \el + 1}} \:
  \sshc{\f}{\el}{\m}{\spin} \:
  \sshc{\wavs}{\el}{0}{\spin}^\cconj \:
  \shfarg{\el}{\m}{\sa}
  \spcend ,
\end{equation}
where we have noted \eqn{\ref{eqn:ssh_wigner}}.  Notice that the
scaling coefficients are a scalar function on the sphere, even when
analysing a spin function \fs\ (\cf\ the wavelet coefficients).
Clearly, the spherical harmonic coefficients of the scaling
coefficients read
\begin{equation}
  \label{eqn:swav_harmonic_coeff}
  \shc{\bigl(\wcoeff^{\swavs}\bigr)}{\el}{\m}
  = 
  \sqrt{\frac{4 \pi}{2 \el + 1}} \:
  \sshc{\f}{\el}{\m}{\spin} \:
  \sshc{\wavs}{\el}{0}{\spin}^\cconj
  \spcend, 
\end{equation}
where
$ \shc{\bigl(\wcoeff^{\swavs}\bigr)}{\el}{\m} =
\innerp{\wcoeff^{\swavs}}{\shf{\el}{\m}}$.

\subsection{Wavelet synthesis}
\label{sec:spin_wavelets:synthesis}

The signal \fs\ can be synthesised exactly from its wavelet and
scaling coefficients by
\begin{equation}
  \label{eqn:wav_synthesis}
  \fs(\sa) 
  = \int_\sphere \dmu{\sa\p} \:
  \scoeff^{\swavs}(\sa\p) \: (\rotarg{\sa\p} \: \swavs)(\sa)
  +
  \sum_{\wscale=\wscalemin}^\wscalemax \int_\sothree \deul{\eul} \:
  \wcoeff^{\swav^\wscale}(\eul) \: (\rotarg{\eul} \: \swav^\wscale)(\sa)
  \spcend ,
\end{equation}
where $\wscalemin$ and $\wscalemax$ are the minimum and maximum
wavelet scales considered, respectively, \ie\
$\wscalemin \leq \wscale \leq \wscalemax$, and are defined explicitly
in \sectn{\ref{sec:spin_wavelets:construction}}.

By decomposing the wavelet and scaling coefficients and functions into
their harmonic expansions, noting the orthogonality of the
spherical harmonics and the Wigner \dmatbig-functions, and also
\eqn{\ref{eqn:ssh_wigner}} and
\eqn{\ref{eqn:spherical_harmonic_coeff_rotation}},
\eqn{\ref{eqn:wav_synthesis}} may be written
\begin{equation}
  \label{eqn:wav_synthesis_harmonic}
  \fs(\sa) 
  =
  \sumlm \biggl[
  \sqrt{\frac{4 \pi}{2 \el + 1}} \:  
  \shc{\bigl(\wcoeff^{\swavs}\bigr)}{\el}{\m} \:
  \sshc{\wavs}{\el}{0}{\spin} 
  +
  \sum_{\wscale=\wscalemin}^\wscalemax \sumn
  \wigc{\bigl(\wcoeff^{\swav^{(\wscale)}}\bigr)}{\el}{\m}{\n} \:
  \sshc{\wav}{\el}{\n}{\spin}^{(\wscale)}
   \biggr ] \sshfarg{\el}{\m}{\sa}{\spin}
  \spcend .
\end{equation}
As we elaborate further in \sectn{\ref{sec:computation:algorithm}},
the inverse wavelet transform of \eqn{\ref{eqn:wav_synthesis}} may
thus be computed via a forward Wigner transform.
Noting \eqn{\ref{eqn:wav_harmonic_coeff}} and
\eqn{\ref{eqn:swav_harmonic_coeff}}, it is clear that \fs\ can only by
synthesised from its wavelet and scaling coefficients through
\eqn{\ref{eqn:wav_synthesis}} if the admissibility condition given by
the following resolution of the identity holds:
\begin{equation}
  \label{eqn:admissibility}
  \frac{4\pi}{2\el+1}   
  \bigl\vert \sshc{\wavs}{\el}{0}{\spin} \bigr\vert^2 + 
  \frac{8\pi^2}{2\el+1} 
  \sum_{\wscale=\wscalemin}^\wscalemax
  \sumn \bigl\vert \sshc{\wav}{\el}{\n}{\spin}^{(\wscale)} \bigr\vert^2 = 1 
  \spcend ,  
  \quad \forall\el 
  \spcend .
\end{equation}

\subsection{Wavelet construction}
\label{sec:spin_wavelets:construction}

We have specified that wavelets should be well-localised in both the
spatial and harmonic domains and satisfy the admissibility condition
of \eqn{\ref{eqn:admissibility}} but we have not yet defined an
explicit construction.  We are now in a position to construct wavelets
that satisfy these properties.  We follow the design of scalar
scale-discretised wavelets \cite{wiaux:2007:sdw,
  leistedt:s2let_axisym, mcewen:2013:waveletsxv,
  mcewen:s2let_localisation}, which are constructed in harmonic
space.  However, rather than construct wavelets in the space of scalar
spherical harmonics, we construct them in the space of spin spherical
harmonics.

Wavelets are defined in spin spherical harmonic space in the
factorised form:
\begin{equation}
  \label{eqn:wav_factorized}
  \sshc{\wav}{\el}{\m}{\spin}^{(\wscale)} \equiv 
  \sqrt{\frac{2\el+1}{8\pi^2}} \:
  \wavker^{(\wscale)}(\el) \: \sshc{\wavsteer}{\el}{\m}{\spin}
  \spcend,
\end{equation}
in order to control their angular and directional localisation
separately, respectively through the kernel
$\wavker^{(\wscale)} \in \ltwo(\reals^{+})$ and directionality
component ${}_\spin\wavsteer \in \ltwo(\sphere)$, with harmonic coefficients
$\sshc{\wavsteer}{\el}{\m}{\spin} = \innerp{{}_\spin\wavsteer}{\sshf{\el}{\m}{\spin}}$.
Without loss of generality, the directionality component is
normalised to impose
\begin{equation}
  \label{eqn:directionality_normalisation}
  \summ \vert \sshc{\wavsteer}{\el}{\m}{\spin} \vert^2 = 1
  \spcend, 
\end{equation}
for all values of \el\ for which $\shc{\wavsteer}{\el}{\m}$ are
non-zero for at least one value of \m.  The angular localisation
properties of the wavelet $\swav^{(\wscale)}$ are then controlled by
the kernel $\wavker^{(\wscale)}$, while the directionality component
$\wavsteer$ controls the directional properties of the wavelet (\ie\
the behaviour of the wavelet with respect to the azimuthal variable
\sab, when centred on the North pole).  

The kernel $\wavker^{(\wscale)}(t)$ is a positive real function, with
argument $t \in \reals^{+}$, although $\wavker^{(\wscale)}(t)$ is 
evaluated only at natural arguments $t=\el$ in
\eqn{\ref{eqn:wav_factorized}}.  The kernel controls the angular
localisation of the wavelet and is constructed to be a smooth function
with compact support, as follows.  Consider the infinitely
differentiable Schwartz function with compact support $t \in
[\dilparam^{-1}, 1]$, for dilation parameter $\dilparam \in \realsnz$,
$\dilparam>1$:
\begin{equation}
  s_\dilparam(t) 
  \equiv s\biggl( \frac{2\dilparam}{\dilparam-1} (t-\dilparam^{-1})-1\biggr)
  \spcend,
\end{equation}
where
\begin{equation}
  s(t) \equiv \Biggl\{ \begin{array}{ll} 
  \ 
  {\rm exp}\bigl(-(1-t^2)^{-1}\bigr), & t\in (-1,1) \\ \  0, & t \notin (-1,1)\end{array} \spcend.
\end{equation}
Define the smoothly decreasing function $k_\dilparam$ by
\begin{equation}
  k_\dilparam(t) \equiv \frac{\int_{t}^1\frac{{\rm d}t^\prime}{t^\prime}s_\dilparam^2(t^\prime)}{\int_{\dilparam^{-1}}^1\frac{{\rm d}t^\prime}{t^\prime}s_\dilparam^2(t^\prime)}, 
\end{equation}
which is unity for $t<\dilparam^{-1}$, zero for $t>1$, and is smoothly
decreasing from unity to zero for $t \in [\dilparam^{-1},1]$.  Define
the wavelet kernel generating function by
\begin{equation}
  \wavker_\dilparam(t) \equiv \sqrt{ k_\dilparam(\dilparam^{-1} t) - k_\dilparam(t) }
  \spcend,
\end{equation}
which has compact support $t \in [\dilparam^{-1}, \dilparam]$ and
reaches a peak of unity at $t=1$.  The scale-discretised wavelet
kernel for scale \wscale\ is then defined by\footnote{We adopt the
  \wscale\ indexing convention followed by
  \cite{leistedt:s2let_axisym} where increasing \wscale\ corresponds
  to smaller angular scales but higher frequency content, which
  differs to the convention adopted in \cite{wiaux:2007:sdw,
    mcewen:2013:waveletsxv, mcewen:s2let_localisation}.}
\begin{equation}
  \wavker^{(\wscale)}(\el) \equiv  \wavker_\dilparam(\dilparam^{-\wscale} \el)
  \spcend,
\end{equation}
which has compact support on
$\el \in \bigl[\floor{\dilparam^{\wscale-1}},
\ceil{\dilparam^{\wscale+1}} \bigr]$,
where $\floor{\cdot}$ and $\ceil{\cdot}$ are the floor and ceiling
functions respectively, and reaches a peak of unity at
$\dilparam^{\wscale}$.
%

The directionality component \wavsteer\ is constructed to carefully
control the directional localisation of the wavelet.  An azimuthal
band-limit \nmax\ is imposed such that $\shc{\wavsteer}{\el}{\m}=0$,
$\forall \el,\m$ with $\vert \m \vert \geq \mmax$.  The directionality
component is defined by imposing a specific form for the directional
auto-correlation of the wavelet, as outlined in \cite{wiaux:2007:sdw,
  mcewen:s2let_localisation} and presented in detail in
\cite{mcewen:s2let_localisation}.  The resulting directionality
component reads
\begin{equation}
  \sshc{\wavsteer}{\el}{\m}{\spin}
  = \eta \: \upsilon
  \sqrt{\frac{1}{2^{p}} \biggl( {{p}\atop{(p-m)/2}} \biggr)}
  \spcend,
\end{equation}
where $\eta = 1$ for $\mmax-1$ even and $\eta = \img$ for $\mmax-1$
odd, $\upsilon = [1 - (-1)^{\mmax+\m}]/2$ and
$p = \min \{ \mmax-1, \el - [1 + (-1)^{\mmax+\el}]/2 \}$.  We have
imposed the symmetries ${}_\spin \wavsteer(\saa, -\sab) = (-1)^{N-1}
\: {}_\spin\wavsteer(\saa, \sab)$ and ${}_\spin \wavsteer(\saa,
\sab+\pi) = (-1)^{N-1} \: {}_\spin\wavsteer(\saa, \sab)$. 

Scaling functions are required to probe the low-frequency content of
the signal of interest not probed by the wavelets and are thus defined
by
\begin{equation}
  \sshc{\wavs}{\el}{\m}{\spin} 
  \equiv 
  \sqrt{\frac{2\el+1}{4\pi}}
  \sqrt{ k_\dilparam(\dilparam^{-\wscalemin} \el)} \: \kron{\m}{0}
  \spcend.
\end{equation}

The maximum wavelet scale $\wscalemax$ is set to ensure the wavelets
reach the band-limit \elmax\ of the signal of interest, yielding
$\wscalemax = \ceil{\log_\dilparam(\elmax-1)}$.  The minimum wavelet
scale $\wscalemin$ may be freely chosen, provided
$0 \leq \wscalemin < \wscalemax$.  For $\wscalemin=0$, the wavelets probe the entire frequency content of the signal of interest
except its mean, which is incorporated in the scaling coefficients.

\begin{figure}
\centering
\includegraphics[width=0.75\textwidth]{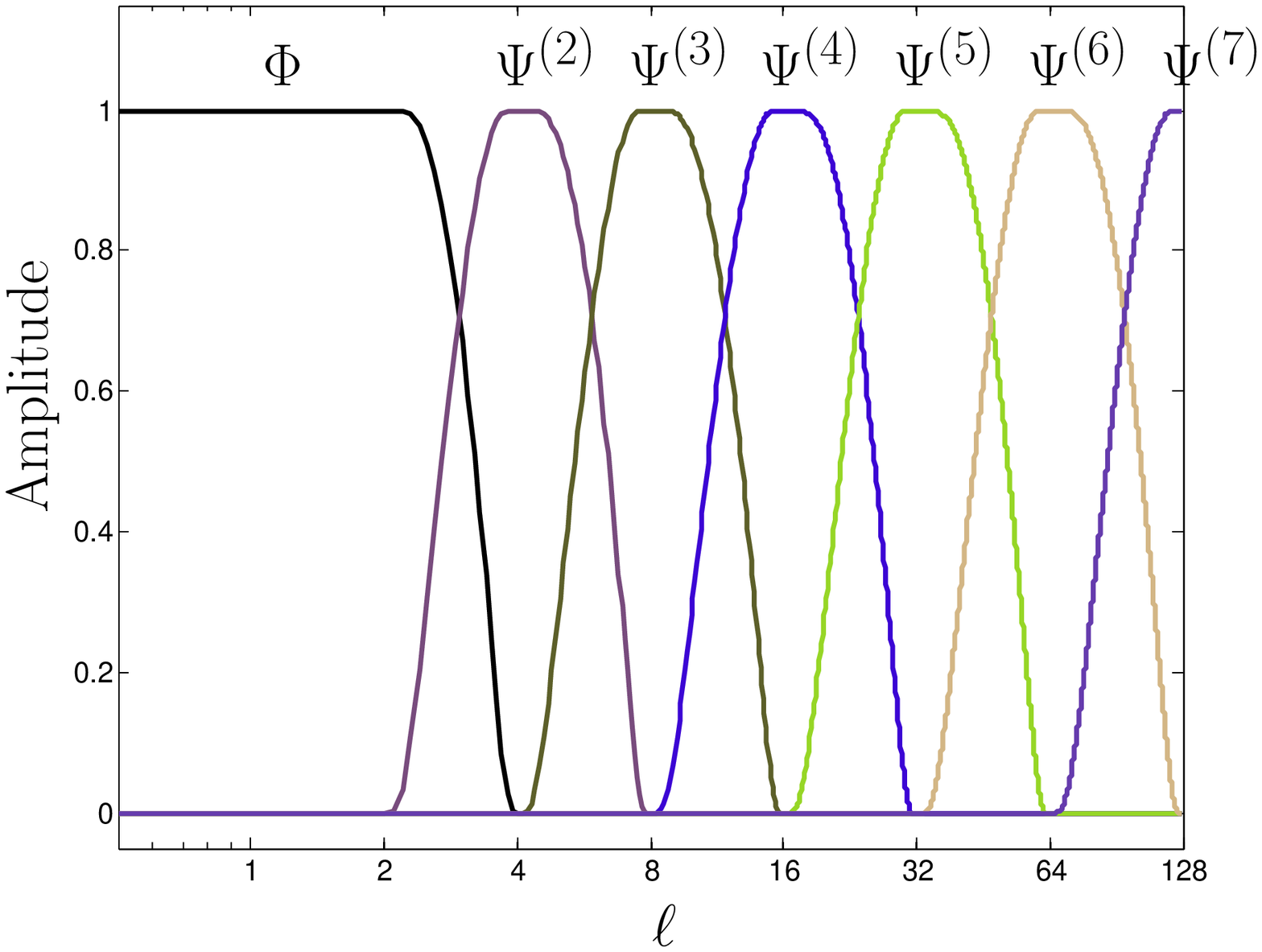}
\caption{Scale-discretised wavelet tiling in spin spherical harmonic
  space (\mbox{$\elmax=128$}, $\dilparam=2$, $\wscalemin=2$,
  $\wscalemax=7$).}
\label{fig:tiling}
\end{figure}

\begin{figure}
    \centering
    \subfigure[Real$\bigl\{{}_2\wav^{(\wscale=1)}\bigr\}$]{\includegraphics[viewport = 95 328 177 410,
      clip=true, width=0.28\columnwidth]{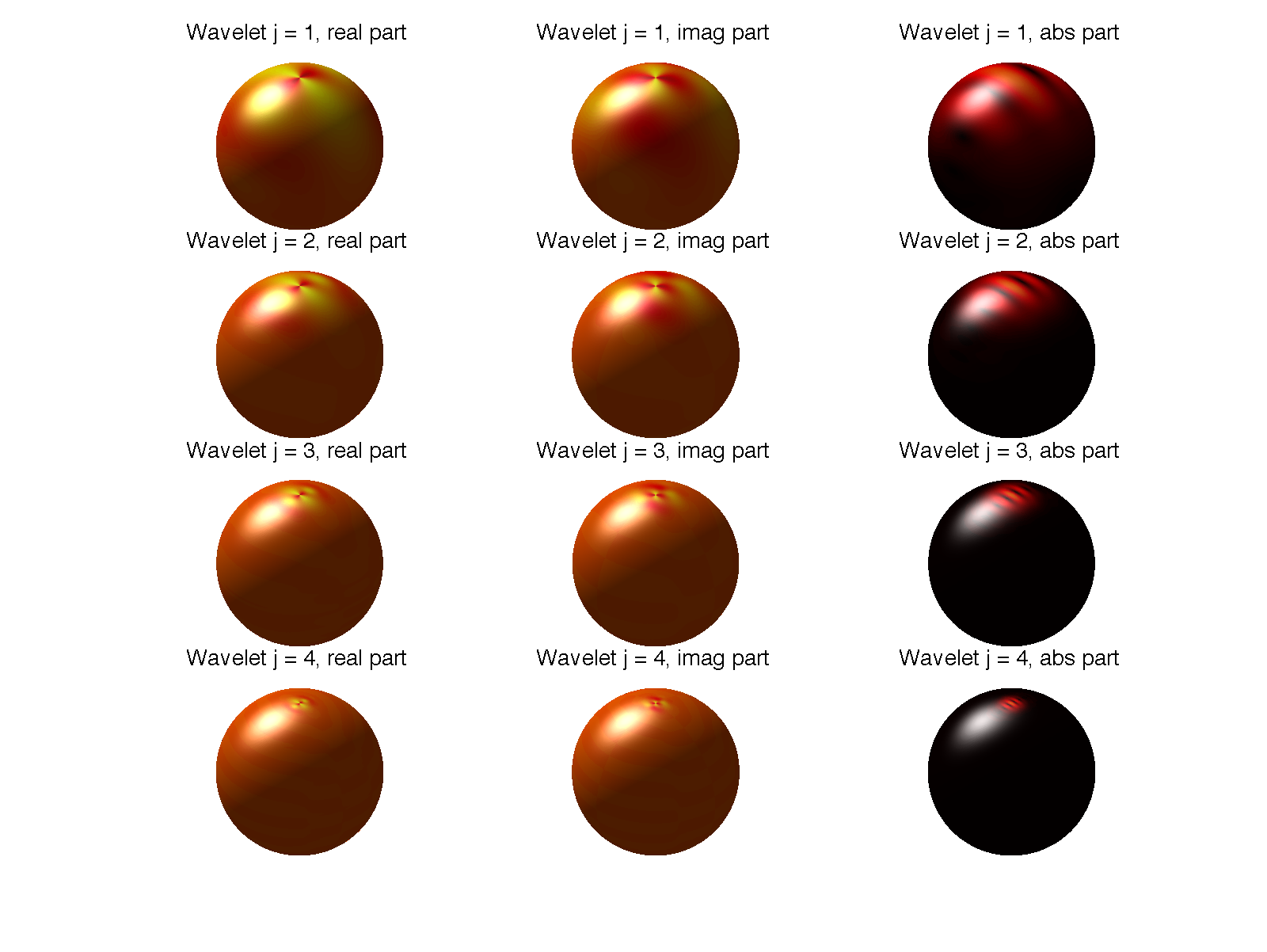}}
    \subfigure[Imag$\bigl\{{}_2\wav^{(\wscale=1)}\bigr\}$]{\includegraphics[viewport = 257 328 339 410,
      clip=true, width=0.28\columnwidth]{Figures/s2let_demo5_Spin2_N5_L512_B2_Jmin2_wav_hot}}
    \subfigure[Abs$\bigl\{{}_2\wav^{(\wscale=1)}\bigr\}$]{\includegraphics[viewport = 420 328 502 410,
      clip=true, width=0.28\columnwidth]{Figures/s2let_demo5_Spin2_N5_L512_B2_Jmin2_wav_hot}}\\
    \subfigure[Real$\bigl\{{}_2\wav^{(\wscale=2)}\bigr\}$]{\includegraphics[viewport = 95 233 177 315,
      clip=true, width=0.28\columnwidth]{Figures/s2let_demo5_Spin2_N5_L512_B2_Jmin2_wav_hot}}
    \subfigure[Imag$\bigl\{{}_2\wav^{(\wscale=2)}\bigr\}$]{\includegraphics[viewport = 257 233 339 315,
      clip=true, width=0.28\columnwidth]{Figures/s2let_demo5_Spin2_N5_L512_B2_Jmin2_wav_hot}}
    \subfigure[Abs$\bigl\{{}_2\wav^{(\wscale=2)}\bigr\}$]{\includegraphics[viewport = 420 233 502 315,
      clip=true, width=0.28\columnwidth]{Figures/s2let_demo5_Spin2_N5_L512_B2_Jmin2_wav_hot}}\\
%
    \caption{Spin scale-discretised wavelets on the sphere ($\spin=2$,
      $\dilparam=2$, $\mmax=5$).  Notice that the absolute value of
      the wavelet is directional, allowing it to probe signal content
      localised not only in scale and position but also orientation.}
    \label{fig:wavelets}
  \end{figure}

By this construction the wavelets and scaling functions
tile the harmonic line $\el$, as illustrated in
\fig{\ref{fig:tiling}}, while satisfying the admissibility condition
of \eqn{\ref{eqn:admissibility}}.  Wavelets are well-localised
simultaneously in the spatial domain, both in position and
orientation, and the harmonic domain.  It is shown in
\cite{mcewen:s2let_localisation} that scale-discretised wavelets
exhibit excellent concentration properties, both in the scalar setting
and in the spin setting presented in this article.  Examples of spin
scale-discretised wavelets are plotted in \fig{\ref{fig:wavelets}}.
Notice that the absolute value of the spin scale-discretised wavelets
is directional.

\subsection{Steerability}

A function on the sphere is steerable if an azimuthal rotation of the
function can be written as a linear combination of weighted basis
functions.  The steerability property of scalar scale-discretised
wavelets extends directly to the spin setting.  By imposing an
azimuthal band-limit \nmax\ on the directionality component such that
$\sshc{\wavsteer}{\el}{\m}{\spin}=0$, $\forall \el,\m$ with
$\vert \m \vert \geq \nmax$, we recover wavelets that are steerable
\cite{wiaux:2007:sdw}.  Moreover, if $T \in \naturals$
of the harmonic coefficients $\sshc{\wavsteer}{\el}{\m}{\spin}$ are
non-zero for a given $\m$ for at least one $\el$, then the number of
basis functions $M \in \naturals$ required to steer the wavelet
directionality component satisfies $M \geq T$ and the optimal number
$M=T$ can be chosen.  Furthermore, if ${}_\spin \wavsteer$ exhibits an
azimuthal band-limit, then it can be steered using basis functions
that are in fact rotations of itself:
\begin{equation}
  \label{eqn:steerability_wav}
  (\rot_{(0,0,\eulc)} {}_\spin \wavsteer)(\sa) = 
  \sum_{\eulci=0}^{M-1} \steerinterp(\eulc - \eulciang) \: 
  (\rot_{(0,0,\eulciang)}{}_\spin\wavsteer)(\sa)
  \spcend,
\end{equation}
where $g \in \naturals$.
The rotation angles $\eulciang \in [0, 2\pi)$ and interpolating
function $\steerinterp \in \ltwo(\reals)$ are defined subsequently.
Note that the interpolating function is independent of the
directionality component ${}_\spin \wavsteer$ of the wavelet.  Due to the
linearity of the wavelet transform, the steerability property is
transferred to the wavelet coefficients themselves, yielding
\begin{equation}
  \label{eqn:steerability_wcoeff}
  \wcoeff^{\swav^{(\wscale)}}(\euls) = 
  \sum_{\eulci=0}^{M-1} \steerinterp(\eulc - \eulciang) \:
  \wcoeff^{\swav^{(\wscale)}}(\eula,\eulb,\eulciang)  
  \spcend.
\end{equation}

Steerability may be proved, and the interpolating functions defined
explicitly, by considering the harmonic expansion of
\eqn{\ref{eqn:steerability_wav}}.  It follows that the
Fourier coefficients of the interpolating function are given by
$\steerinterp_{\m} = 1/M$ and the symmetries of ${}_\spin \wavsteer$
outlined in \sectn{\ref{sec:spin_wavelets:construction}} may be
exploited to optimise the number of basis functions to $M=\mmax$, with
equiangular sampling $\eulciang = \eulci \pi / M$.  The derivation is
entirely analogous to the scalar setting \cite{wiaux:2007:sdw,
  mcewen:s2let_localisation}.

\begin{figure}
  \centering
\mbox{
  \raisebox{11mm}{\footnotesize $\steerinterp_{\eulc_0}$}\hspace*{-1.5mm}
  \includegraphics[viewport= 155 80 440 370, clip=true, width=0.21\columnwidth]{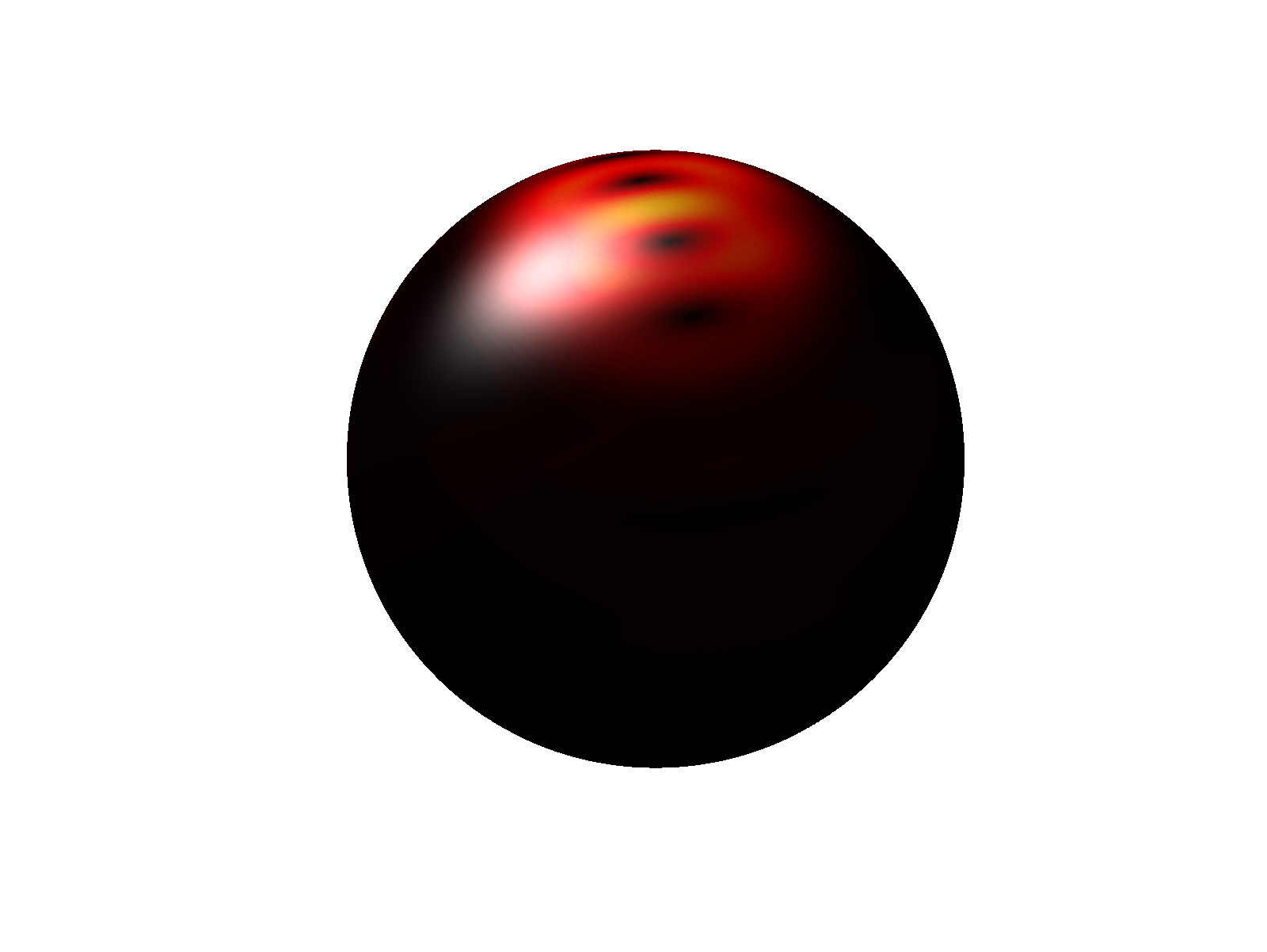}
  \hspace*{-3mm}
  \raisebox{11mm}{\footnotesize $+\steerinterp_{\eulc_1}$}
  \hspace*{-3mm}
  \includegraphics[viewport= 155 80 440 370, clip=true, width=0.21\columnwidth]{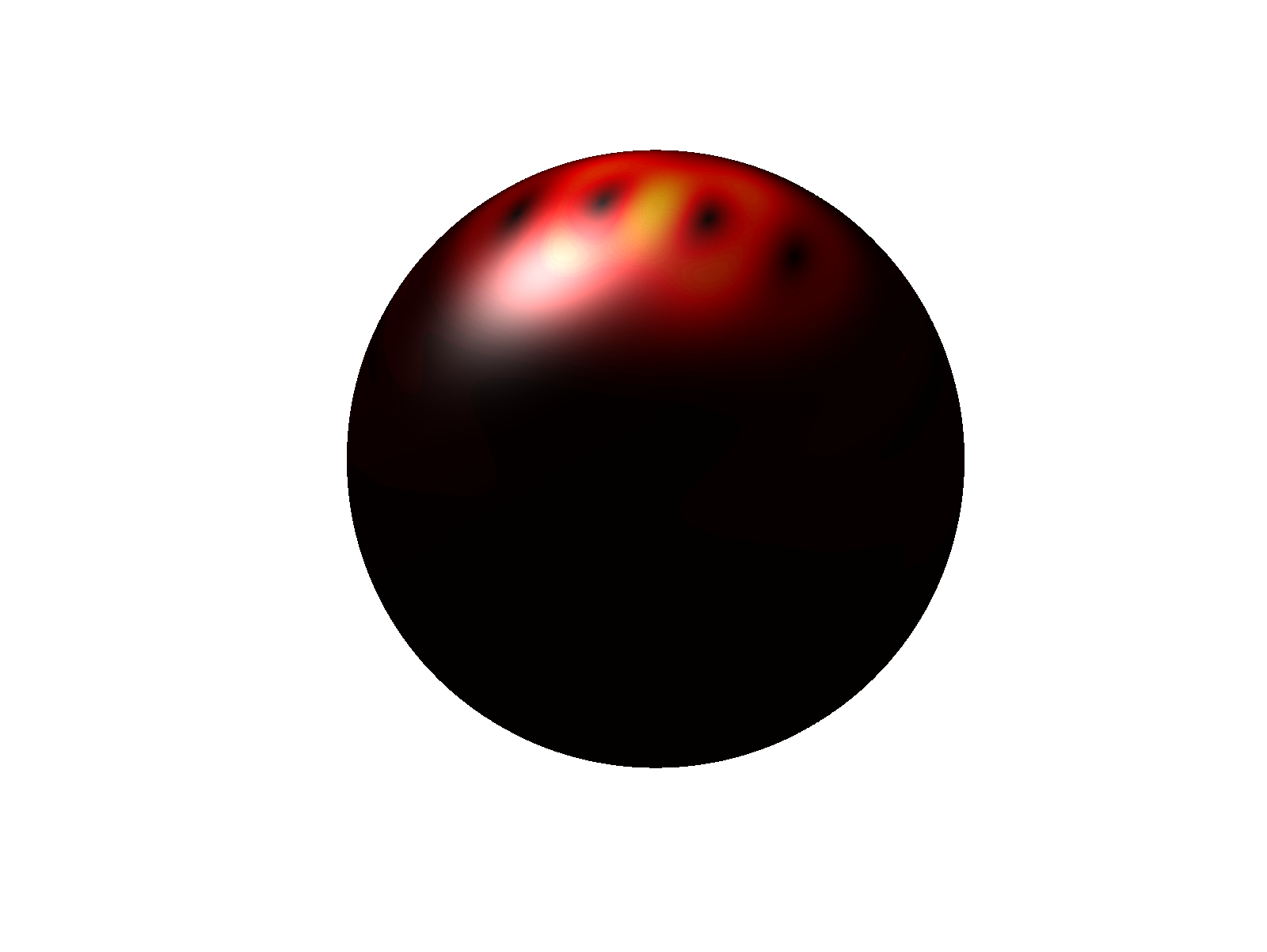}
  \hspace*{-3mm}
  \raisebox{11mm}{\footnotesize $+\steerinterp_{\eulc_2}$}
  \hspace*{-3mm}
  \includegraphics[viewport= 155 80 440 370, clip=true, width=0.21\columnwidth]{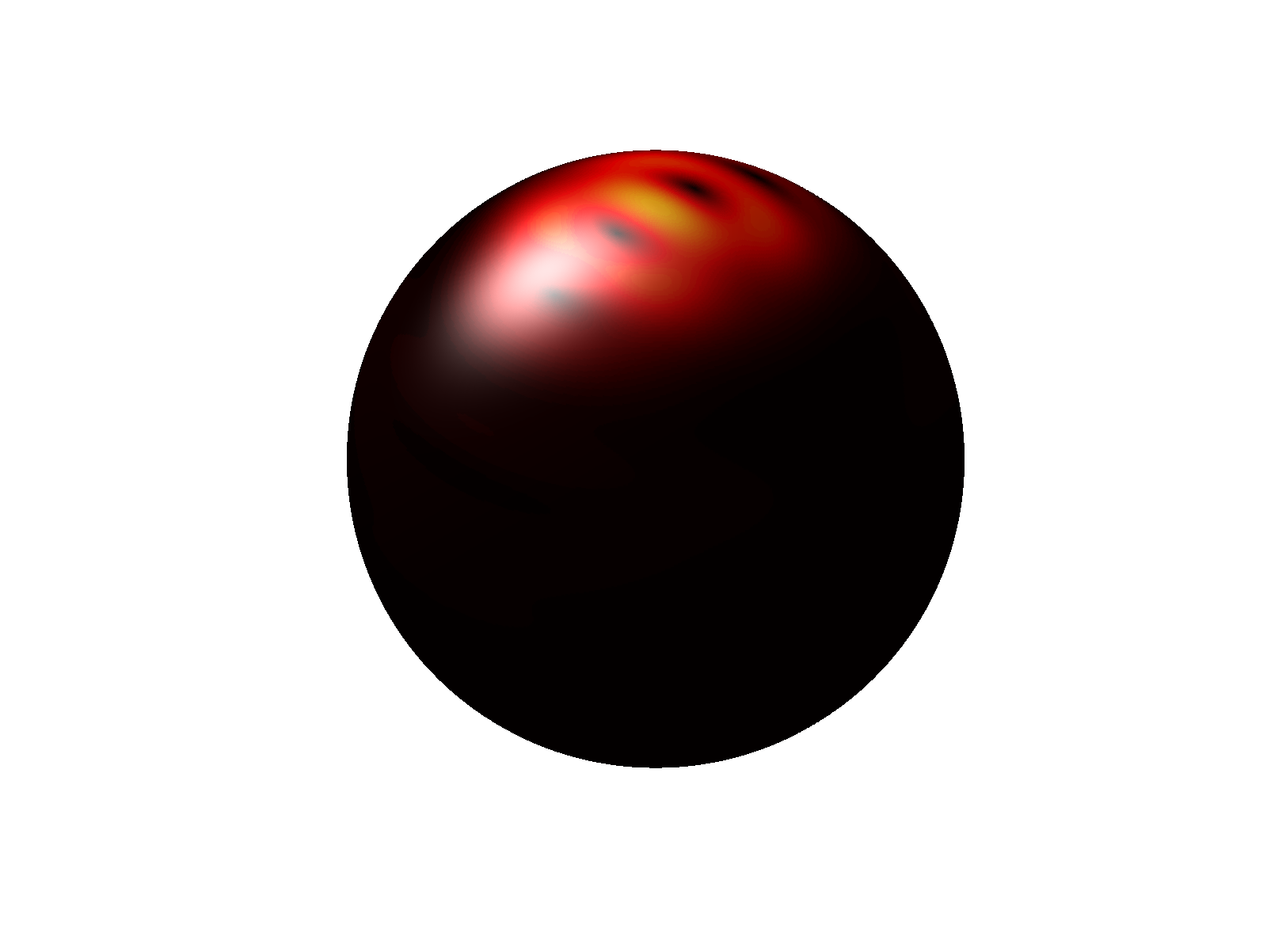}
}
  \: \raisebox{11mm}{\small $=$}\:
  \includegraphics[viewport= 155 80 440 370, clip=true, width=0.21\columnwidth]{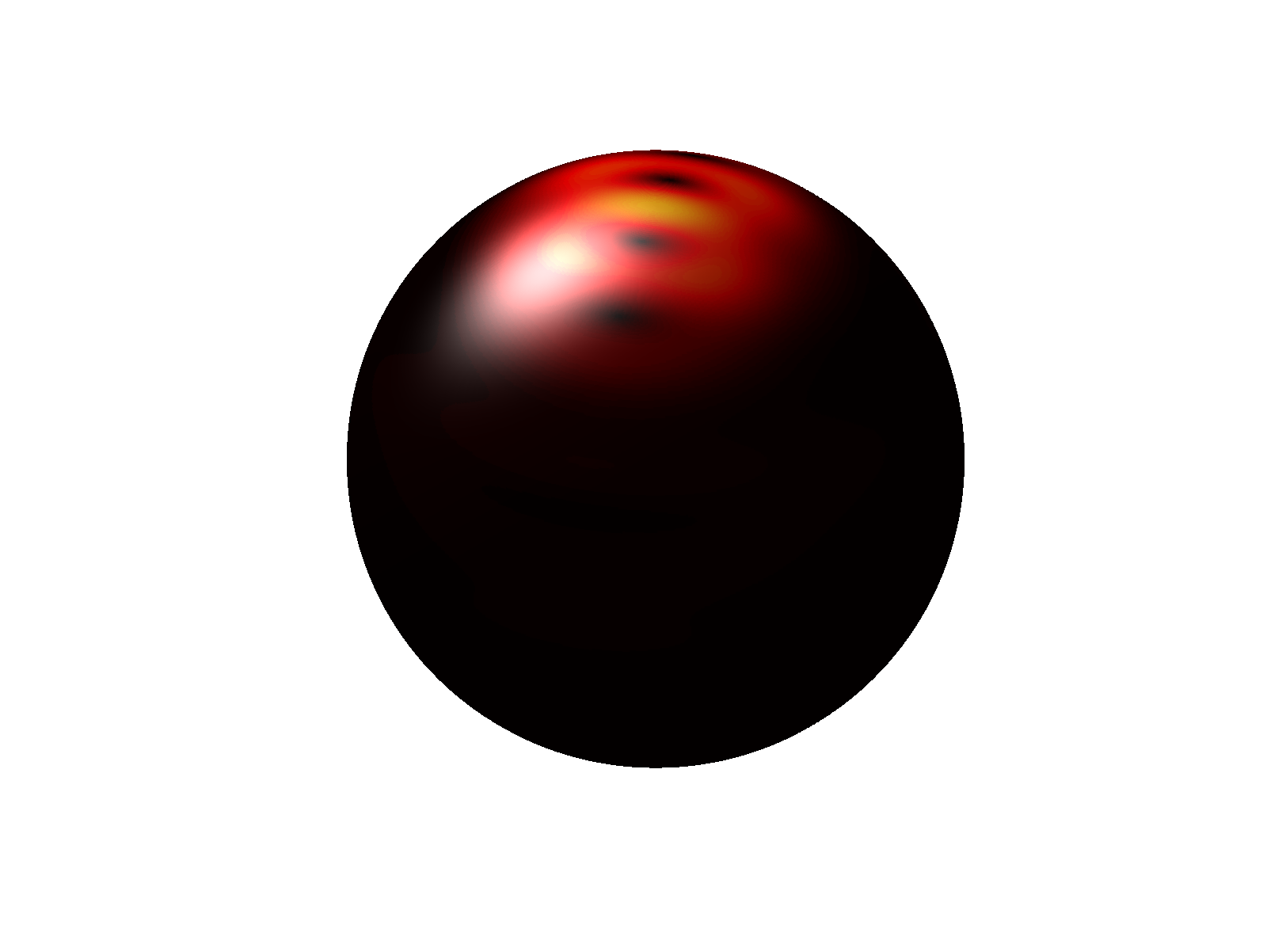}
  \vspace{-2mm}
  \caption{Absolute value of basis and steered spin wavelets
    ($\spin=2$, $\dilparam=2$, $\mmax=3$, $\wscale=3$).  The wavelet
    can be steered to any continuous orientation $\eulc$ by taking
    weighted sums of its three basis wavelets.  }
  \label{fig:steerability}
\end{figure}

A steered wavelet for $\mmax=3$ and its basis wavelets, given by
rotated versions of the wavelet, are plotted in
\fig{\ref{fig:steerability}}.  The wavelet can be steered to any
continuous orientation $\eulc$ by taking weighted sums of its three
basis wavelets.

\subsection{Parseval frame}

Spin scale-discretised wavelets on the sphere satisfy the following tight
frame property:
\begin{equation}  
  A \: \|  \fs \|^2
  \leq 
  \int_\sphere \dmu{\sa} 
  \bigl\vert \innerp{\fs}{\rotarg{\sa} \: \swavs} \bigr\vert^2 
  +
  \sum_{\wscale=\wscalemin}^\wscalemax \int_\sothree \deul{\eul}
  \bigl\vert \innerp{\fs}{\rotarg{\eul}\:\swav^{(\wscale)}} \bigr\vert^2  
  \leq 
  B \: \| \fs \|^2
  \spcend ,
  \label{eqn:frame_property}
\end{equation}
with $A=B\in\realsnz$, for any band-limited $\fs \in \ltwo(\sphere)$,
and where $\| \cdot \|^2 = \innerp{\cdot}{\cdot}$. We adopt a
shorthand integral notation in \eqn{\ref{eqn:frame_property}},
although by appealing to exact quadrature rules \cite{mcewen:fssht,
  mcewen:so3} these integrals may be replaced by finite sums.

This property may be proved by considering the term between inequalities
in \eqn{\ref{eqn:frame_property}}, substituting harmonic expansions
of all terms
and noting the orthogonality of the spin spherical
harmonics and the Wigner $\dmatbig$-functions.  By the wavelet
admissibility condition \eqn{\ref{eqn:admissibility}} it follows that
this term
reduces to $ \| \fs \|^2$.
Spin scale discretised wavelets hence constitute a Parseval frame on
the sphere, where $A=B=1$.

\subsection{Connection to scalar wavelet transform}
\label{sec:spin_wavelets:scalar_connection}

For spin $\spin = \pm 2$ signals, a connection can be made between the
spin wavelet transform of ${}_{2}(\qpiu)$ and scalar wavelet
transforms of E- and B-mode signals, which are not directly
observable.
First, consider the wavelet coefficients of the observable
${}_2(\qpiu)$ signal computed by a \emph{spin} wavelet transform:
$\wcoeff^{{}_2 \wav^{(\wscale)}}_{{}_2(\qpiu)}(\eul) \equiv
\innerp{{}_2(\qpiu)}{\rotarg{\eul}\:{}_2 \wav^{(\wscale)}}$.
Second, consider the wavelet coefficients of the unobservable
$\emodetilde$ and $\bmodetilde$ signals computed by a \emph{scalar}
wavelet transform:
$\wcoeff^{{}_0 \tilde{\wav}^\wscale}_{\emodetilde}(\eul) \equiv
\innerp{\emodetilde}{\rotarg{\eul}\:{}_0 \tilde{\wav}^\wscale}$
and
$ \wcoeff^{{}_0 \tilde{\wav}^\wscale}_{\tilde{B}}(\eul) \equiv
\innerp{\bmodetilde}{\rotarg{\eul}\:{}_0 \tilde{\wav}^\wscale}$.
If the wavelet used in the scalar wavelet transform is a spin lowered
version of the wavelet used in the spin wavelet transform, \ie\
${}_0 \tilde{\wav}^\wscale = \bar{\eth}^2 {}_2 \wav^\wscale$, then the
wavelet coefficients of $\emodetilde$ and $\bmodetilde$ are simply related
to the wavelet coefficients of ${}_2 (\qpiu)$ by
$\wcoeff^{{}_0 \tilde{\wav}^\wscale}_{\emodetilde}(\eul) = - {\rm Re}
\bigl[ \wcoeff^{{}_2 \wav^\wscale}_{{}_2 (\qpiu)} (\eul) \bigr]$
and $\wcoeff^{{}_0 \tilde{\wav}^\wscale}_{\bmodetilde}(\eul) = - {\rm
  Im}\bigl[\wcoeff^{{}_2 \wav^\wscale}_{{}_2 (\qpiu)} (\eul)\bigr]$,
respectively.  
Similar connections for spin-2 signals exist for standard and
mixed needlets \cite{geller:2008, geller:2010:sw, geller:2010}.  

\section{Exact and efficient computation}
\label{sec:computation}

The scale-discretised wavelet transform of band-limited spin signals
on the sphere can be computed exactly and efficiently by appealing to
sampling theorems on the sphere \cite{driscoll:1994, mcewen:fssht} and
rotation group \cite{mcewen:so3} and corresponding fast transforms.
By exploiting these sampling theorems we develop a new fast algorithm,
which is theoretically exact, to compute the scale-discretised wavelet
transform to very high band-limits.  Moreover, our algorithm requires
approximately half as many wavelet coefficients to capture the full
information content of band-limited signals compared to alternative
approaches \cite{mcewen:2013:waveletsxv}.  We then discuss and
evaluate the numerical implementation of this algorithm.

\subsection{Fast algorithm}
\label{sec:computation:algorithm}

As noted previously, forward and inverse wavelet transforms may be
computed via inverse and forward Wigner transforms, respectively.  We
therefore defer the majority of the computation of the
scale-discretised wavelet transform to fast algorithms that we 
developed recently to compute Wigner transforms \cite{mcewen:so3}.

In a recent article \cite{mcewen:so3} we developed a novel sampling
theorem for signals defined on the rotation group, such as directional
wavelet coefficients, by associating the rotation group with the
three-torus through a periodic extension.  All of the information
content of a band-limited signal can be captured in $4\elmax^3$
samples, reducing the number of required samples by a factor of two
compared to other \mbox{equi\-angular} sampling theorems 
\cite[\eg][]{driscoll:1994}.  Moreover, we developed fast algorithms to
compute Wigner transforms to very high band-limits, which are also
theoretically exact.  We exploit these developments to compute spin
scale-discretised wavelet transforms exactly and efficiently.  The
efficient nature of our sampling theorem on the rotation group
\cite{mcewen:so3} means that only half as many wavelet
coefficients must be computed compared to alternative approaches
\cite{mcewen:2013:waveletsxv}, while still capturing the full
information content of the signal under analysis.  

The forward wavelet transform of \eqn{\ref{eqn:wav_analysis}} for a
given scale $\wscale$ can be represented by the inverse Wigner
transform of \eqn{\ref{eqn:wav_analysis_harmonic}}, with Wigner
coefficients given by \eqn{\ref{eqn:wav_harmonic_coeff}}.  For each
scale $\wscale$, the complexity of computing
\eqn{\ref{eqn:wav_harmonic_coeff}} is $\order(\mmax \elmax^2)$, while
the complexity of computing \eqn{\ref{eqn:wav_analysis_harmonic}} is
reduced from the naive case of $\order(\mmax^2 \elmax^4)$ to
$\order(\mmax \elmax^3)$ by our fast Wigner transform
\cite{mcewen:so3}.  Recall that $\mmax$ is the azimuthal band-limit of
the wavelet.  Scaling coefficients can be computed by
\eqn{\ref{eqn:swav_harmonic_coeff}} at $\order(\elmax^2)$, followed
by an inverse spherical harmonic transform at $\order(\elmax^3)$ by
exploiting our novel sampling theorem and corresponding fast 
spherical harmonic transforms \cite{mcewen:fssht}.  For a given scale
$\wscale$, computing the forward wavelet transform is dominated by the
inverse Wigner transform and thus scales as $\order(\mmax \elmax^3)$.

The inverse wavelet transform of \eqn{\ref{eqn:wav_synthesis}} can be
computed via forward Wigner transforms.  It is apparent from
\eqn{\ref{eqn:wav_synthesis_harmonic}} that the spherical harmonic
coefficients of \fs\ can be recovered by
\begin{equation}
  \sshc{\f}{\el}{\m}{\spin}
  =   
  \sqrt{\frac{4 \pi}{2 \el + 1}} \:  
  \shc{\bigl(\wcoeff^{\swavs}\bigr)}{\el}{\m} \:
  \sshc{\wavs}{\el}{0}{\spin} 
  +
  \sum_{\wscale=\wscalemin}^\wscalemax \sumn
  \wigc{\bigl(\wcoeff^{\swav^{(\wscale)}}\bigr)}{\el}{\m}{\n} \:
  \sshc{\wav}{\el}{\n}{\spin}^{(\wscale)}
  \spcend .
\end{equation}
Firstly, Wigner coefficients
$\wigc{\bigl(\wcoeff^{\swav^{(\wscale)}}\bigr)}{\el}{\m}{\n}$ must be
computed by a forward Wigner transform of the wavelet coefficients
$\wcoeff^{\swav^{(\wscale)}}(\eul)$.  By exploiting our novel sampling
theorem on the rotation group and fast Wigner transform,
which connects the rotation group \sothree\ to the three-torus through a periodic extension and appeals to FFTs on the three-torus,
the
computation of Wigner coefficients for a given scale $\wscale$ can be
reduced from the naive case of $\order(\mmax^2 \elmax^4)$ to
$\order(\mmax \elmax^3)$ (for further details see \cite{mcewen:so3}).
The contribution to $\sshc{\f}{\el}{\m}{\spin}$ for a given scale
$\wscale$ can then be computed at $\order(\mmax \elmax^2)$.  The
contribution from the scaling coefficients can be computed at
$\order(\elmax^2)$ after a spherical harmonic transform at
$\order(\elmax^3)$ by exploiting our fast spherical harmonic transforms
\cite{mcewen:fssht}.
For a given scale $\wscale$, computing the contribution to synthesise
the original signal from its wavelet coefficients is dominated by the
forward Wigner transform and thus scales as $\order(\mmax \elmax^3)$.

We have so far considered individual wavelet scales only.  Naively,
when including all
$\wscalemax_{\rm tot} = \wscalemax - \wscalemin + 1$ wavelet scales
$\wscale$, both the forward and inverse wavelet transforms scale as
\mbox{$\order( \wscalemax_{\rm tot} \mmax \elmax^3)$}.  However, since
the wavelets themselves have compact support in harmonic space (see
\fig{\ref{fig:tiling}}) it is not necessary to perform all wavelet
transforms at the full band-limit \elmax.  Furthermore, the lower
harmonic support of the wavelets for large scales (smaller $\wscale$) can also be exploited to increase
computational efficiency.  A multi-resolution algorithm is constructed
following these optimisations, where the computation is dominated by the
largest two scales.  Hence, the overall complexity of computing both
forward and inverse wavelet transforms, including all scales, is
effectively $\order(\nmax \elmax^3)$. In addition, for real signals we exploit their conjugate symmetry to further reduce computational and memory requirements by a factor of two.

\subsection{Implementation}

We have implemented the exact and efficient algorithm described
previously to compute spin scale-discretised wavelet transforms on the
sphere in the existing \stwoletcode\ code
\cite{leistedt:s2let_axisym}, which has also been extended to support
directional wavelets.  The core algorithms of \stwoletcode\ are
implemented in C, while Matlab, Python and IDL interfaces are
also provided.  Consequently, \stwoletcode\ is able to handle very
large harmonic band-limits, corresponding to data-sets containing tens
of millions of pixels.
\stwoletcode\footnote{\url{http://www.s2let.org}} is publicly
available, and relies on the
\sshtcode\footnote{\url{http://www.spinsht.org}} code
\cite{mcewen:fssht} to compute spherical harmonic transforms, the
\sothreecode\footnote{\url{http://www.sothree.org}} code
\cite{mcewen:so3} to compute Wigner transforms and the
\fftwcode\footnote{\url{http://www.fftw.org}} code to compute Fourier
transforms.  Note that it also supports the analysis of data on the
sphere defined in the common
\healpix\footnote{\url{http://healpix.jpl.nasa.gov}}
\cite{gorski:2005} format.

\subsection{Numerical evaluation}

We evaluate the numerical accuracy and computation time of the spin
scale-discretised wavelet transform implemented in the \stwoletcode\
code.  We simulate band-limited test signals on the sphere defined by
uniformly random spherical harmonic coefficients
$\sshc{\f}{\el}{\m}{\spin}$ with real and imaginary parts distributed
in the interval $[-1,1]$.  We then compute an inverse spherical
harmonic transform to recover a band-limited test signal on the
sphere.  A forward wavelet transform is then performed, followed by an
inverse transform to synthesise the original signal from its wavelet
coefficients.  Ten simulated signals are considered for band-limits
from $\elmax=32$ to $\elmax=2048$.  All numerical experiments are
performed on a standard desktop with a 3.5$\:$GHz Intel Core i7
processor and 32$\:$GB of RAM.

\subsubsection{Numerical accuracy}

\begin{figure}
\centering
\includegraphics[width=.75\textwidth, trim = 0.0cm 0.5cm 0cm 0.0cm, clip]{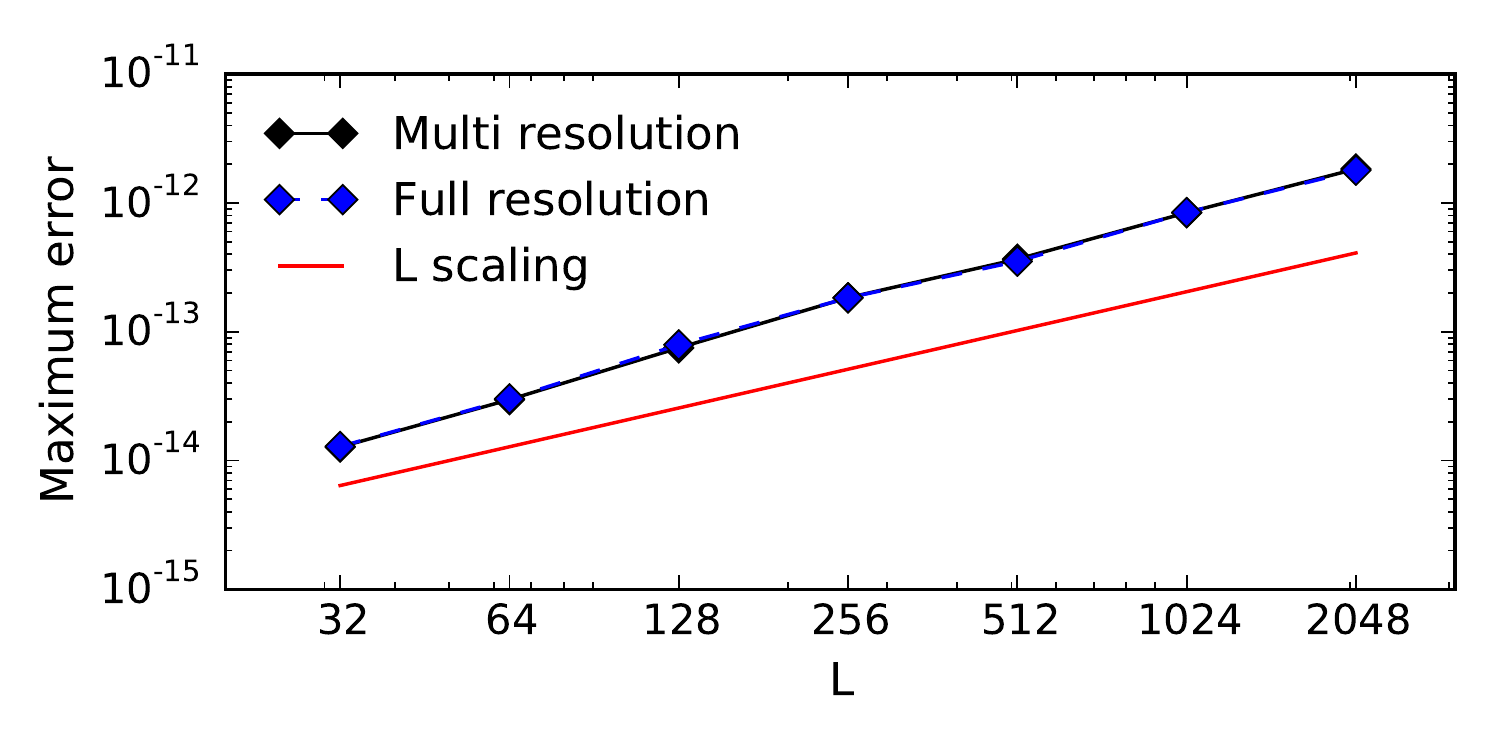}
\caption{Numerical accuracy of a round-trip wavelet transform
  ($\spin=2$, $\mmax = 5$, $\dilparam = 2$).  Accuracy is close to
  machine precision and found empirically to scale as
  $\order(\elmax)$.}
\label{fig:accuracy}
\end{figure}

Numerical accuracy of a round-trip wavelet transform is
measured by the maximum absolute error between the spherical harmonic
coefficients of the original test signal $\fslm^{\rm o}$ and the
recomputed values $\fslm^{\rm r}$, \ie\
$
  \epsilon = \mathop{\rm max}_{\el,\m} \:
  \bigl | \fslm^{\rm r} - \fslm^{\rm o} \bigr |
$.
Results of the numerical accuracy tests, averaged over ten random test
signals, are plotted in \fig{\ref{fig:accuracy}}.  We plot results for
a spin $\spin=2$ signal, although the accuracy for different spin
numbers is identical.  The numerical accuracy of the round-trip
transform is close to machine precision and found empirically to scale
as $\order(\elmax)$.

\subsubsection{Computation time}

\begin{figure}
\centering
\includegraphics[width=.75\textwidth, trim = 0.0cm 0.5cm 0cm 0.0cm,
clip]{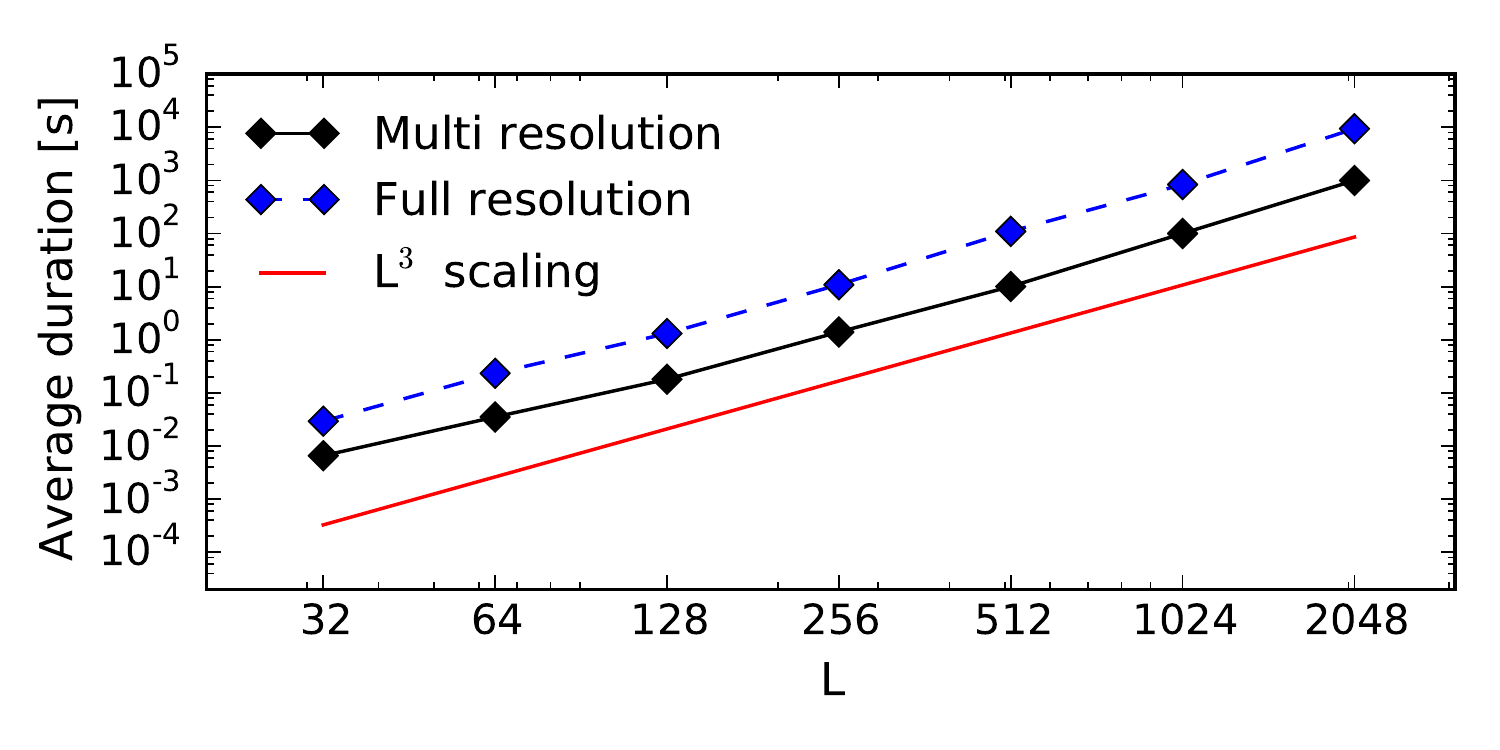}
\caption{Computation time of a round-trip wavelet transform
  ($\spin=2$, $\mmax = 5$, $\dilparam = 2$).  For constant $\mmax$,
  computation time is found empirically to scale as
  $\order(\elmax^3)$, as expected. Furthermore, the multiresolution
  transform is approximately an order of magnitude faster than the
  full-resolution transform.}
\label{fig:timing}
\end{figure}

Computation time is measured by the round-trip computation time taken
to perform a forward and inverse wavelet transform.  Results of the
computation time tests, averaged over ten random test signals, are
plotted in \fig{\ref{fig:timing}}.  We plot results for a spin
$\spin=2$ signal, although the computation time for different spin
numbers is identical since the spin number is simply a parameter of
the transform (rather than applied through spin lowering/raising operators).  For constant $\mmax$, computation time is found
empirically to scale as $\order(\elmax^3)$, as predicted in
\sectn{\ref{sec:computation:algorithm}}.

\section{Applications}
\label{sec:applications}

We discuss applications of our directional spin scale-discretised wavelet
analysis on the sphere in this section.  Firstly, we illustrate their use in a
simple denoising example.  Secondly, we give brief overviews of recent
applications of spin scale-discretised wavelets for cosmological analyses.

\subsection{Illustration}

We consider linearly polarised radiation and
construct the spin-2 signal $ x={}_2 (\qpiu) \in \ltwo(\sphere)$ from the
$Q$ and $U$ Stokes parameters, as described in
\sectn{\ref{sec:harmonic_analysis:spin2}}. The observed signal
$ y \in \ltwo(\sphere)$ is contaminated with zero-mean
Gaussian white noise on the sphere $ n\in \ltwo(\sphere)$, with
variance 
$
  \mathbb{E}\bigl( \vert  {n}_{\ell m} \vert^2 \bigr) \ = \ \sigma^2, \quad \forall \ell,\ m, $
and thus reads $ y =  x +  n$, where all signals are
band-limited at \elmax.

To assess the fidelity of noisy signals, we consider the
signal-to-noise ratio (SNR) defined by 
\begin{equation}
  \textrm{SNR}(y) \equiv 10 \log_{10} \frac{ \|  x \|_2^2 }{ \|  y -  x \|_2^2 }
\spcend ,
\end{equation}
where signal energies are calculated by
\begin{equation}
  \|  x \|_2^2 \equiv \innerp{ x}{ x} 
  = \int_{\sphere} \dmu{\sa}  \vert  x(\sa) \vert^2 
  = \sumlmbl \vert {}x_{\ell m} \vert^2
  \spcend .
\end{equation}

We estimate $ x$ by denoising $ y$, to recover the denoised
signal $ \hat{x} \in L^2(S^2)$,
taking advantage of the fact that most natural signals of interest
(\eg, generated by physical processes) have their energy concentrated
in a small number of wavelet coefficients. By contrast, the energy of
the noise is spread over all wavelet scales. Since the wavelet
transform is linear, the wavelet coefficients of the observed signal
for the $j$-th scale are simply given by the sum of signal and noise
contributions: 
\begin{equation}
  Y^{{}_2\wav^{(\wscale)}}(\eul) = X^{{{}_2\wav^{(\wscale)}}}(\eul) +  N^{{}_2\wav^{(\wscale)}}(\eul)
  \spcend,
\end{equation}
where upper-case variables denote wavelet coefficients of the
corresponding signal  denoted by lower-case variables, \ie\
\mbox{$Y^{{{}_2\wav^{(\wscale)}}} \equiv y \convdir
  {{}_2\wav^{(\wscale)}}$}. Due, again, to the linearity of the wavelet
transform, the wavelet coefficients of the noise are also
zero-mean and Gaussian, with variance 
\begin{equation}
  \mathbb{E} \bigl( \vert{N}^{{{}_2\wav^{(\wscale)}} }(\eul)\vert^2 \bigr) 
  =  \sigma^2 \sum_{\ell n} \bigl\vert  {}_2\wav^{(\wscale)}_{\ell n} \bigr\vert^2 
  \equiv  \bigl( \sigma^{(\wscale)} \bigr)^2
  \spcend . 
\end{equation}
Thus, a simple denoising strategy is to hard-threshold the wavelet
coefficients $Y^{{{}_2\wav^{(\wscale)}}}$ with a threshold
\mbox{$t^{(\wscale)} = 3\sigma^{(\wscale)}$}. This yields denoised wavelet
coefficients 
\begin{equation}
	\hat{X} ^{{}_2\wav^{(\wscale)}}(\eul) = \left\{\begin{array}{ll}
		0 ,   & \textrm{if } Y^{{}_2\wav^{(\wscale)}}(\eul)< t^{(\wscale)} \\
		Y^{{}_2\wav^{(\wscale)}}(\eul),    &\textrm{otherwise}
		\end{array}\right. . \label{threshold}
\end{equation}
Scaling coefficients are not thresholded: we simply take
$\hat{X}^{{}_2\wavs}=Y^{{}_2\wavs}$. The denoised signal $ \hat{x} $ is then
reconstructed from its wavelet coefficients
$\hat{X}^{{}_2\wav^{(\wscale)}}$ and scaling coefficients $\hat{X}^{{}_2\wavs}$
by an inverse spin wavelet transform.

\begin{figure}
\centering
\includegraphics[width=.75\textwidth, trim = 3cm 0.9cm 0.5cm 0.9cm, clip]{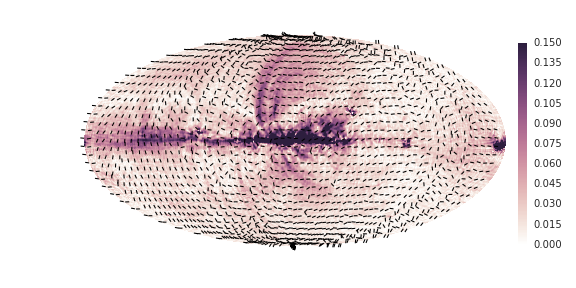}
\vspace{-3mm}
\caption{Full-sky polarised synchrotron emission inferred from WMAP
  data \cite{gold:2008}, plotted using a Mollweide projection, in
  units of $\mu$K.  The colour scale of the plot shows the amplitude
  of the spin-2 signal ${}_2 (\qpiu)$, while the direction of the
  headless vectors show its phase.  These data provide the test signal
  used for the denoising illustration.  }
\label{fig:illustration_data}
\end{figure}

\begin{figure}
\centering
\subfigure[$\vert {}_2 (\qpiu) \vert$]{\includegraphics[width=\textwidth, trim = 15.5cm 0.9cm 0.9cm 0.9cm, clip]{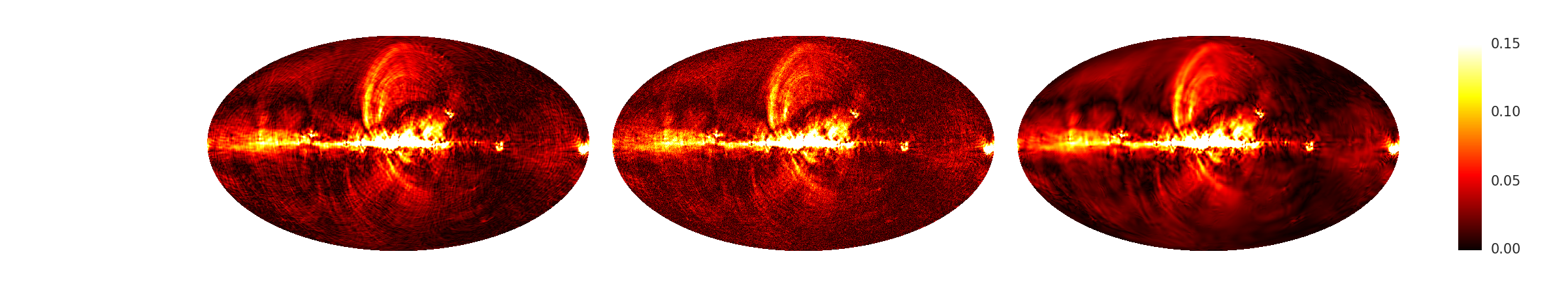}}
\vspace{-2mm}
\subfigure[$\angle\: {}_2 (\qpiu) $]{\includegraphics[width=\textwidth, trim = 15.5cm 0.9cm 0.9cm 0.9cm, clip]{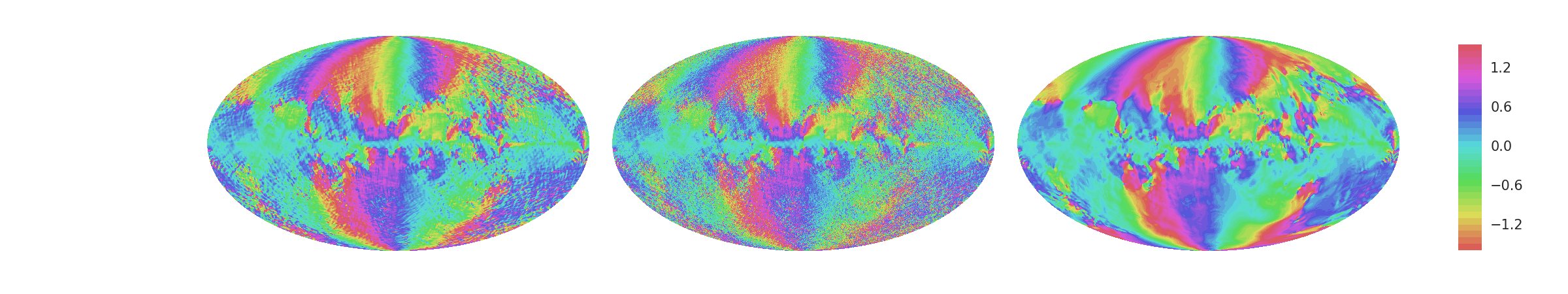}}
\vspace{-1mm}
\caption{Wavelet denoising of the full-sky polarised synchrotron
  emission shown in \fig{\ref{fig:illustration_data}}.  The amplitude
  of the spin-2 signal ${}_2 (\qpiu)$ is shown in panel (a) in units
  of $\mu$K, while its phase is shown in panel (b).  Noisy data is
  plotted on the left and denoised data recovered by a simple
  hard-thresholding strategy is plotted on the right.  Noisy data is
  constructed with an SNR of 11$\:$dB, while denoised data is
  recovered with an SNR of 18$\:$dB.}
\label{fig:illustration_denoising}
\end{figure}

This simple denoising strategy is demonstrated using the full-sky
polarised synchrotron emission that is inferred from WMAP data
\cite{gold:2008} and shown in \fig{\ref{fig:illustration_data}}, which
is used as the input test signal $ x$. This test signal is highly
structured and generated by real astrophysical
processes. Consequently, it is a good example signal for a denoising
illustration, even though it already contains some noise. The test
signal is corrupted with noise to yield the observed signal $ y$,
such that $\textrm{SNR}( y)=11\:$dB. By applying the denoising
algorithm described above (with parameters $\elmax=512$, $\lambda=2$,
$N=4$, and $J_0=0$) a denoised signal $ \hat{x}$ is recovered with
\mbox{$\textrm{SNR}(d)=18\:$dB}.  Noisy and denoised data are plotted
in \fig{\ref{fig:illustration_denoising}}.  Repeating the denoising
using axisymmetric spin scale-discretised wavelets (by setting $N=1$)
degrades the fidelity of the denoised signal by $\sim1$dB. In this
illustration, where a very simple hard-thresholding algorithm
is adopted, the denoising is surprisingly effective given that the
input signal is taken from real astrophysical data and already contains
residual noise.

\subsection{Cosmological applications}

Just as directional scalar wavelets have found widespread use in cosmological studies, particularly for the analysis of CMB temperature observations, we anticipate directional spin wavelets will be similarly useful in cosmology.  In particular, CMB polarisation \cite{zaldarriaga:1997} and cosmic shear due to weak gravitational lensing \cite{heavens:2003} give rise to spin-2 signals on the celestial sphere.  In a number of recent articles some of the authors of this article have already applied the spin wavelet framework constructed here to cosmological problems, as described below (these are only a small selection of cosmological examples where directional spin wavelets are likely to be useful and are not an extensive set of examples).

\subsubsection{CMB component separation}

In \cite{rogers:s2let_ilc_pol} the Spin-SILC algorithm is developed to
remove foreground contamination from multi-frequency polarised CMB
observations.  Spin-SILC is applied to \emph{Planck} data to recover a clean estimate
of the CMB polarisation signal. This approach acts coherently on the observed
spin quantity (the Stokes parameters $\qpmiu$), unlike alternatives that
consider $\stokesq$ and $\stokesu$ (or $E$ and $B$) independently.

\subsubsection{Weak gravitational lensing}

In \cite{leistedt:flaglets_spin} our spin wavelet framework is extended
from the sphere to the 3D ball.  Moreover, we express the theory of 3D weak
gravitational lensing in directional spin wavelet space and construct
wavelet covariance estimators to relate data to cosmological theory.  This
approach allows one to handle effectively, and simultaneously, complicated sky
coverage and uncertainties associated with the physical modelling of small
scales.

\subsubsection{E/B reconstruction}

In \cite{leistedt:ebsep} we exploit the connection between spin wavelet
transforms of observed spin-2 signals and scalar wavelet transforms of the
associated E- and B-mode quantities (see
\sectn{\ref{sec:spin_wavelets:scalar_connection}}) to recover E- and B-modes
from partial-sky cosmological observations.  The E/B reconstruction problem
arises when analysing both CMB polarisation and weak gravitational lensing.
Performing E/B separation in wavelet space supports the application of scale-
and orientation-dependent masking schemes.  Furthermore, we
derive wavelet space pure-mode estimators, which rely on spin-1 wavelet
transforms (also afforded by our spin wavelet framework since our construction
supports arbitrary spin), reducing leakage due to the partial-sky setting by
over an order of magnitude.

\section{Conclusions}
\label{sec:conclusions}

The directional spin scale-discretised wavelet framework constructed
in this article is the only wavelet framework defined natively on the
sphere that is able to probe the directional intensity of spin
signals.  The wavelet transform can thus be used to probe signal
content not only in scale and location but also orientation.  This
additional degree of freedom improves the sensitivity of wavelet
analyses in practical applications, as demonstrated already for scalar
signals in numerous analyses of the temperature anisotropies of the
\cmb\ \cite[\eg][]{mcewen:2005:ng, mcewen:2006:isw, mcewen:2007:isw2}.  In addition, our directional
spin scale-discretised wavelet framework supports the exact synthesis
of a signal from its wavelet coefficients, in theory and in practice,
and is applicable to signals of arbitrary spin, while the wavelets
themselves constitute a Parseval frame and are steerable, implying
wavelet coefficients for any continuous orientation can be computed
from a finite number of fixed orientations.  Furthermore, we present
exact and efficient algorithms to compute the forward and inverse
wavelet transform for very large data-sets containing tens of millions
of samples on the sphere.  We require only half as many wavelet
coefficients to capture the full information content of the signal
compared to alternative approaches.  Finally, we highlight the
applicability of spin directional wavelets by a simple illustration of
an application to denoising linearily polarised radiation observed on
the celestial sphere, which is a spin $\spin = \pm 2$ signal.

\bibliographystyle{elsarticle-harv}
\bibliography{bib_myname,bib_journal_names_long,bib}

%








\end{document}